# Top performers and top journals:
# Persistent concentration in scientific publishing

**Marek Kwiek**
Center for Public Policy Studies, Adam Mickiewicz University, Poznan, Poland
kwiekm@amu.edu.pl, ORCID: orcid.org/0000-0001-7953-1063

**Wojciech Roszka**
(1) Poznan University of Economics and Business, Poznan, Poland
(2) Center for Public Policy Studies, Adam Mickiewicz University, Poznan, Poland
wojciech.roszka@ue.poznan.pl, ORCID: orcid.org/0000-0003-4383-3259

## Abstract

In this research, we analyze the relationship between publishing productivity and access to highly prestigious journals, treating publishing in top journals as a stratification mechanism selecting publishing elites. We study $N$ = 144,314 Polish scientists publishing for 30 years (1992–2021) and their *Nart* = 433,546 unique research articles published in the period. Using bibliometric data from Scopus, we compare the scientists belonging to the top productivity decile (the upper 10%, termed top performers) and the remaining population of scientists (90%) by discipline and period (five six-year periods). We measure the share of publications in prestigious segments of journals, with particular reference to the 90th–99th percentiles, and we use nonlinear journal prestige-normalized productivity. Our results indicate that access to top journals (defined as the top 10% of journals indexed in Scopus) is powerfully and permanently concentrated in the group of top performers in all disciplines and periods studied. The differences between top performers and the other scientists are primarily of a qualitative nature: they are seen almost exclusively at the top of the journal hierarchy rather than in its bottom or middle segments. Our logistic regression models indicate the complementarity of quantity and quality: publishing intensity increases the probability of membership in the elite segment of top performers, especially when it is coupled with publishing in prestigious journals. Our results suggest that top journals function as selection gates to academic careers and that they function as durable mechanisms of elite reproduction in science.

# 1. Introduction

Inequalities in publishing productivity have been documented for about a century (Cole & Cole, 1973; de Solla Price, 1963; Lotka, 1926; Merton 1968). The distribution of publications in all disciplines is extremely steep: a small number of scientists account for a disproportionately high part of all knowledge production (Allison & Stewart, 1974; Kwiek, 2016). Sociological analyses of stratification in science have shown that the concentration in science does not refer only to publication quantity but also to citations, access to resources, and recognition and awards (Cole & Cole, 1973). A key element in explaining the permanence of these inequalities over time is cumulative advantage: in a system of rewards and communication in science, earlier successes increase the probability of later successes, and recognition is distributed in an asymmetric and self-reinforcing manner (Allison & Stewart, 1974; Allison et al., 1982). In other words, success (in science) breeds success, and the rich (in science) get richer.

But productivity concentration is not the only mechanism that feeds academic stratification. What is equally important – and is often critical to successful academic careers – is the concentration of scholarly journals through which knowledge is published because journal prestige functions as an institutional signal of quality and visibility (Heckman & Moktan, 2020; Leahey, 2007). From the perspective of citation behavior, the impact is related not only to article quality: citations too reflect an author's reputation, their national and international visibility, and the prestige of publication channels. Citations, in a word, are sensitive to journal hierarchies (Bornmann & Daniel, 2008).

In the last two decades, publishing in highly prestigious journals (historically, always important in science as a provider of international recognition) has become a central reference point in many research assessment systems. Expectations of publishing in top journals are standard among national research funding agencies (Lyytinen et al., 2007), and they are an absolute must for the European Research Council (Rodríguez-Navarro & Brito, 2019). In most extreme examples, this logic takes the form of career gatekeeping through a narrow set of elite journals. In economics, for instance, as an extensive analysis by Heckman and Moktan shows, the location of a publication in the so-called "top-five" journals is powerfully linked to positive tenure and promotion decisions (Heckman & Moktan, 2020). Many disciplines have their narrowly defined lists of top journals in which the vast majority of scientists are never able to publish or in which they publish just once (as the first author has shown for top journals in higher education research, about 80% of all authors in the six top published only one paper in the 20 years studied, and only 3.3% had five publications or more; Kwiek, 2021). Different disciplines have different publishing cultures that define what is regarded as a top journal.

In biology, medicine, chemistry, and physics, the system of top journals is global, relatively stable, and based on Impact Factor (from WoS) or CiteScore (from Scopus) measures; in the social sciences, hierarchies of journals exist, but they are more contestable, with economics being the most hierarchized discipline; and in the humanities, the hierarchies of journals are more dispersed, less formalized, and less unequivocal (Hammarfelt, 2017). In technical and engineering fields, hierarchies of journals are often linked to hierarchies of conferences. Finally, in applied and interdisciplinary fields, there are different competitive lists of top journals (e.g., FT50 in management). For biology and medicine, top journals include *Nature* and *Science*; for economics, the top five include *The American Economic Review* and *Econometrica* (Heckman & Moktan, 2000); for management, top journals include *Management Review* and *Administrative Science Quarterly;* and for higher education research, they include *Higher Education* and *Studies in Higher Education.*

Not all institutions are involved to the same extent in the prestige game based on publication in top journals (Kwiek, 2021). Elite institutions in national systems tend to focus more on top journals compared with average institutions: what matters for research-intensive universities may not matter at all for comprehensive teaching-focused universities. There are also substantial cross-national differences in the role of publications in top journals for hiring, promotion, and research funding decisions (see discussions of top journals in different systems in, e.g., Bak & Kim, 2019; Fochler et al., 2016; Lindahl, 2018; Mouritzen & Opstrup, 2020; Shibayama & Baba, 2015; Sutherland, 2018).

Previous studies have often used a simplified separation of productivity (publications) and impact (citations), with productivity defined as the number of publications per scientist (within a unit of time). However, as Abramo and D'Angelo (2014) showed, such an account of productivity is conceptually problematic and leads to misguided assessments of research efficiency at various levels. They propose quality-adjusted measures of productivity that link publishing output to quality and co-authorship (Abramo & D'Angelo, 2014). In practice, this means that in studies of scientific elites, it is not enough to ask, "Who publishes more?" "Who publishes where?" also needs to be asked, and the impact that selective access to prestige channels has on positions in the science system must also be assessed. In this research, we use nonlinear journal prestige-normalized productivity (see explanations in section 3.3. Methods).

The aim of this article is to empirically disentangle two dimensions of stratification in science: publishing productivity and access to the most prestigious journals. Moving beyond simple quantitative measures, we investigate whether the advantage of top performers (the upper 10% of scientists in publishing productivity) over the remaining scientists (90%) is linear and dispersed in the entire hierarchy of journals – or whether it emerges selectively and only in the highest segments of productivity. We pay special attention to the segment of journals from the 90–99 percentile (the upper 10% of journals) in the Scopus dataset – which, in STEM disciplines, is a powerfully selective segment of the global publishing circuit.

We link our analysis of publication elites to classical analyses of inequalities and cumulative advantage in science (Cole & Cole, 1973; de Solla Price, 1963; Lotka, 1926; Merton, 1968). We show that the prestige of the publication channel is a critical dimension of visibility and selection in science (Bornmann & Daniel, 2008; Leahey, 2007). We view the patterns discovered in the context of works that treat top journals as powerfully supporting careers, especially visible in disciplines with strong hierarchies of journals (as in economics; Heckman & Moktan, 2020). We change the productivity perspective from the question of "who publishes more" (quantity) to the question of "who publishes in what channels" (quantity and *ex ante* quality, *ex post* quality being related to citations), and we discuss the consequences of this new perspective for stratification in science.

In empirical terms, we studied 144,314 unique scientists publishing over 30 years (between 1992 and 2021) and their 433,546 unique research articles published in that period and indexed in Scopus. The data were aggregated into five six-year analytical periods and 15 STEM disciplines.

## 2. Literature review

The first theoretical pillar of this research comprises studies on top performers in science, which grow out of a long tradition of analyses showing that publication productivity is structurally unequal (Abramo et al., 2009; Abramo et al., 2013; Abramo et al., 2017; Ioannidis et al., 2014; Rosen, 1981; Xie, 2014). Classical works concluded that the distribution of productivity at the individual level is extremely steep: the majority of scientists publishes little or nothing, and a small minority produces a considerable share of knowledge (de Solla Price, 1963; Lotka, 1926). As Lotka (1926, p. 317) remarked in a paper presenting his "inverse square law" of scientific productivity, "it would be of interest to determine, if possible, the part which men of different calibre contribute to the progress of science." Classical observations about the role of the highly productive tiny minority of scientists in knowledge production laid the foundations for later interpretations of social stratification in science in which knowledge production is hierarchically organized, and recognition and resources (of various types, from funding and top positions in universities to research infrastructure) are allocated unevenly (Cole & Cole, 1973).

In the sociology of science, of special importance is an approach in which inequalities in science result from institutional and cumulative mechanisms. Merton (1968, p. 7) applied the idea of cumulative advantage not only to individuals but also to institutions: "the rich get richer at a rate that makes the poor become relatively poorer. Thus, centers of demonstrated scientific excellence are allocated far larger resources for investigation than centers which have yet to make their mark." Cole and Cole (1973) described science as a stratification system in which prestige, visibility, and access to resources form a feedback loop with productivity. In their account, better positions in the scientific field do not result exclusively from individual differences in talent or effort – mechanisms of selection and the reproduction of hierarchies are involved. These interpretations were supported by, e.g., Allison and Stewart (1974), who showed that differences in productivity can be partly explained by the process of cumulative advantage; however, the accumulative advantage hypothesis is not able to explain productivity differences between the youngest scientists (where the "sacred spark hypothesis" could work: some scientist are not good – they are just superb, from the very beginning; see Stephan, 2012; Stephan & Levin, 1996). Allison and Stewart (1974) argue that cumulative advantage mechanisms should produce inequalities as cohorts of scientists grow older. Allison et al. (1982) developed these logics into a more general theory of inequalities in science explaining that advantages in recognition and resources increase in a self-reinforcing way. Instead of synthetic cohorts, they used true cohorts of scientists. They also suggested that their theory of inequality in science might be generalizable to other areas, e.g., to the distribution of wealth.

The principal theoretical framework for cumulative advantage processes is the Matthew effect, which means that earlier recognition and success increase the probability of later recognition and success (both symbolic and institutional) and that both tend to be concentrated: "eminent scientists get disproportionately great credit for their contributions to science while relatively unknown scientists tend to get disproportionately little credit for comparable contributions" (Merton, 1968, p. 2). Although Merton's focus was on allocating credit and recognition (Merton, 1973), in bibliometric analyses that followed, the Matthew effect was interpreted more widely: as a mechanism that acts through institution prestige, the reputation of the author, and recognition of communication channels, including scholarly journals.

Classical literature provides two arguments important to our research questions. First, top performers (also termed "eminent scientists," "star scientists," or "prolific scientists") are a structural element of the science system (Abramo et al., 2009; Fox & Nikivincze, 2021; Prpić, 1996). And second, inequalities in productivity and recognition are supported by cumulative mechanisms, which may be coupled with selective access to prestigious publication channels.

The second theoretical pillar important to our research is the set of studies on measures of prestige in science. In academic practice, citations and journal prestige often function as proxy indicators of quality. But the literature shows that citations are multidimensional measures encompassing quality, visibility, and reputation. Bornmann and Daniel (2008) indicate that citations are not a simple reflection of research quality: they also depend on journal prestige, author reputations, and the structure of the field.

Journal prestige is a critical conversion mechanism (Latour & Woolgar, 1986): publications in prestigious journals become more visible, they have a higher citation probability, and they offer faster diffusion in the science community, all of which, in the long run, strengthen the author's position in the field (Bak & Kim, 2019; Fochler et al., 2016; Lindahl, 2018). From the perspective of academic careers, visibility functions as a resource that can be exchanged for  awards and opportunities. Leahey (2007) shows that visibility and specialization are significant determinants of academic salaries and that productivity influences careers, to a large extent, through mechanisms of visibility.

If journal prestige increases visibility and strengthens cumulative advantage, then access to top journals may act as an independent mechanism of stratification – not only as a reward for productivity but also as an element of elite selection and reproduction. If "quantitative" productivity (publication counts) is only partly correlated with "qualitative" productivity (publications linked to journals in which they are published), then there is a question of whether publishing elites should be defined through sheer publication numbers or publications in specific (prestigious) journals. We are especially interested in top journals as a potential area where "qualitative productivity" is produced and where cumulative advantage may operate most strongly.

This article draws from our previous research. We have shown elsewhere that the role of top performers in national knowledge production is relatively equal across 11 European systems. Top performers are responsible for about 50% of all publications in those systems – which suggests a structural rather than accidental character of European productivity elites (Kwiek, 2016). Using large-scale survey data, we have developed a catalog of characteristics of top performers (long working hours, long research hours, research orientation rather than teaching orientation, international collaboration, being male). However, survey data could not be linked to publication and journal metadata, so the link between top performers and top journals could not be established (Kwiek, 2018a). In a recent study, we found that the role of top performers in a national science system in the past three decades was unchangeable: the top 1% of Polish scientists were responsible for about 10% and the top 10% of scientists for about 50% of national knowledge production (Kwiek & Roszka, 2024). Top performers emerge from the data as builders of the structure of the scientific field.

Heckman and Moktan (2020) present an institutional argument: under conditions of oversupply of publications and competition for scarce academic positions, the system simplifies the selection to publications located in a tiny set of channels ("top-five" journals) that become formal and informal career criteria. The idea of career gatekeeping through selective access to prestigious channels allows interpretations of empirical difference between top performers and the rest of the scientific population as structural and qualitative, rather than merely quantitative. In this approach, access to top journals is an elite-reproducing mechanism: it increases visibility, enables the accumulation of citations, and strengthens reputation.

Recently, several strands of research have provided additional arguments to strengthen theoretical theses regarding the increasing importance of top journals. Zhang and Yu (2020) analyzed the publishing strategies of 2,982 scientists and determined that publishing in prestigious journals early in one's career is treated as a key quality signal. Hou et al. (2022) studied 1,933 business scholars and showed that early publications in top journals are powerful predictors of long-term career success. Additionally, early bloomers in science are more likely to attract top scientists as collaborators or have opportunities in prestigious institutions. Kraus and colleagues (2023), analyzing the career factors of 100 prominent researchers, highlight the importance of early career factors for future impact, which is consistent with the intuition that entering prestigious channels early can change career trajectories. They reported that 95% of the prominent researchers studied had at least one of four features in the first five years of their careers: working at a top-25 ranked university, publishing a paper in a top-five ranked journal, publishing most papers in top quartile (high impact) journals, or co-authoring with other researchers prominent in their field.

There is ever more work analyzing institutional pressures and the systemic effects of the increasing role of top journals. Seeber (2024) showed that changes in the publishing business may have an impact on authors' decisions on where to publish and their focus on a narrower pool of journal titles. He observed that individual choices regarding where to publish are relevant not only for individual careers and recognition – they also collectively shape scientific communication. Trueblood and colleagues (2025) described a "misalignment of incentives" in academic publishing suggesting that emphasis on publishing in prestigious journals may generate undesirable behaviors among academics, constrain their innovativeness, and strengthen their research conformism.

Finally, literature on the differences between groups of authors formed on the basis of productivity points to permanent stratification linked to top journals. The first author has described (Kwiek, 2021) the concentration of authorship in elite journals and the advantage of frequent publishers have over infrequent ones (in top journals in higher education research in the past 30 years). The empirical results strengthen a view of top journals as a publishing segment with high entry barriers and strong reproduction mechanisms. Romero-Silva and colleagues (2025) discussed the dominance of leading business schools in top journals, and Balzer and Benlahlou (2025) analyzed prestige bias in citations, which is consistent with the argument that cumulative advantage is embedded in prestige channels.

There are two hypotheses explaining the grip of top journals on individual scientists (Fochler et al., 2016; Lindahl, 2018), institutions (Mouritzen & Opstrup, 2020), and national systems (Franzoni et al., 2011). One is the prestige-maximization model (see Kwiek, 2021), located within broader theories of academic capitalism and resource dependence (Slaughter & Leslie, 1997), which explains how universities and scientists work together for the similar goal of prestige maximization. The other is principal-agent theory, which explains how publishing in prestigious journals aligns the interests of individual scientists (as agents) with those of their institutions and research-sponsoring organizations, including national governments (as principals; Braun & Guston, 2003; Kivistö, 2008; van der Meulen, 1998).

According to the prestige-maximization model, research universities, their departments, and individual scientists act as "prestige maximizers" (Melguizo & Strober, 2007; Taylor et al., 2016). Prestige received from publications in top journals can be converted into research grants; and institutions, departments, and scientists modify their behavior, including publishing patterns, accordingly, in competing for external resources (Taylor et al., 2016). Publishing in elite journals became especially important following the development of research assessment systems (van Dalen & Henkens, 2005; Whitley & Glaser, 2007). Research funding from prestigious granting agencies is another avenue of prestige maximization.

Principal-agent theory can be applied to better understand the role of top journals for governments. The theory was initially used in studies of corporations (Pratt & Zeckhauser, 1985), but then it was applied to science sectors (Braun & Guston, 2003; Kivistö, 2008; van der Meulen, 1998). In the relationship between research universities and individual scientists as agents and governments as principals, publication in top journals is a key indicator of institutional productivity and a critical component in competition for funding, including the competition within various national excellence initiatives. (The Polish version of a national excellence initiative, the IDUB Program of 2020–2026, was built entirely on Scopus- and WoS-defined indicators of top journals.)

Generally, it is almost impossible for the principal to understand the agent's products (Braun & Guston, 2003, pp. 303–304), that is, research publications, or to assess their impact. At the same time, the principal must ensure that academics produce high-quality research. The label of publication in top journals enables principals at all levels (national, institutional, and departmental) to defend their distribution of rewards in a rational manner. Top journals can also easily serve as a common performance metric in most disciplines (Gomez-Mejia & Balkin, 1992, p. 925).

Principal-agent theory offers a useful way to understand the appeal of top journals to both principals and agents (for whom it is easier to demonstrate their own academic success). Instead of reading thousands of papers in the way peer review suggests, the theory reduces the burden of quality assessment. The number of papers is radically reduced if only publications in top journals are reported (and they can be easily traced across disciplines, academic units, institutions, and countries).

The integration of the classical literature on inequalities in science, the literature on visibility and prestige, and studies on measures of productivity, publishing elites, and top journals leads to a conclusion that research has often treated the relationship between high productivity and high journal prestige as obvious and unidirectional. Meanwhile, contemporary science is ever more strongly organized by hierarchies of journals that may act as mechanisms of elite selection and reproduction.

Our aim is not to confirm that top performers publish much more than others (as we have done; see Kwiek, 2016, 2018; Kwiek & Roszka, 2024) but to confirm that differences between top performers and other scientists are qualitative in nature: they are concentrated mostly in the highest ranks of journals. This study interprets these patterns as durable stratification mechanisms in which selective access to top journals is a key source of long-term advantage.

## 2.1. Research questions and hypotheses

The aims of this study are to understand the relationships between publication productivity and access to prestigious journals and to assess the extent to which publishing in top journals constitutes a stratification mechanism of the publishing elite. In contrast to approaches that focus exclusively on publication numbers, in this research we assume that productivity and the prestige of publication channels are related in a nonlinear and complementary manner, and access to top journals may serve as a selective career gate. In this context, the following research questions and hypotheses were formulated.

### 2.1.1. Research questions

RQ1. Do top performers publish in top journals (90–99 Scopus journal percentile ranks) to a disproportionately higher extent than other scientists, regardless of discipline or period?

RQ2. Is the observed stratification between top performers and other scientists qualitative in nature – that is, do the differences between the two groups emerge primarily in the highest segments of the journal prestige hierarchy?

RQ3. How does the concentration of publications in top journals translate into the concentration of productivity measured by a journal prestige-normalized indicator?

RQ4. Is entry into the group of top performers more strongly associated with publication intensity itself or rather with a combination of a high number of publications and high quality of publication channels?

On the basis of these research questions, the following research hypotheses were formulated.

### 2.1.2. Research hypotheses

H1 (*Prestige selection*)
Top performers achieve a significantly higher share of publications in top journals (journals located in the 90–99 Scopus percentile ranks) than other scientists in all disciplines and periods studied.

H2 (*Qualitative stratification*)
The differences between top performers and the remaining scientists emerge primarily in the highest journal prestige deciles, while in the lower and middle segments of the journal hierarchy, the publication profiles of the classes are relatively similar.

H3 (*Complementarity of quantity and quality*)
The effect of publishing intensity on the probability of membership in the class of top performers increases with the quality of publication channels, indicating a complementary – rather than a supplementary – effect of quantity and quality of scientific production.

H4 (*Structural character of stratification*)
The stratification of access to top journals is structural: it persists over time (30 years) despite increases in publication volume, and it occurs in all disciplines, albeit with varying intensity.

# 3. Dataset, sample, and methods

## 3.1. Dataset

We used a Scopus dataset with authors and publications extracted in November 2022–January 2023. We also used supplementary data provided by the International Center for the Study of Research (ICSR Lab) run by Elsevier, the owner of the Scopus dataset. We used the complete, in terms of coverage, dataset from 1992 to 2022. This period corresponds to the availability of consistent Scopus coverage for Polish-affiliated authors and allows for the analysis of long-term structural patterns rather than short-term fluctuations. We used data from all Polish research-involved institutions (in all sectors, 343 in total) though the vast majority and authors and publications come from the higher education sector. We used separate affiliations in each of the six five-year periods. Our approach was to select the dominating affiliation in the publications' bylines in the period. However, the only institutional variable used was IDUB vs. all other institutions (the IBUB variable characterizing 10 selected research-intensive universities, all participants in the Polish national research excellence program). This binary distinction was adopted to capture broad differences in research intensity of Polish institutions. No differences are made among non-IDUB institutions, so the class includes all remaining universities, institutes of the Polish Academy of Sciences, research institutes, and corporate bodies. Mobility between institutions, as captured by changing affiliations, is marginal in Poland and is noted especially for young scientists who are employed on research grants. As a result, affiliation-based measurement error is unlikely to substantially affect individual productivity estimates within periods.

Scientists assigned by Scopus to any of the 343 institutions were regarded as Polish scientists. We also treated as scientists co-authors who were graduate students (very rare cases), as they were listed as Scopus authors with individual IDs. Individuals (rather than publications) were the unit of analyses. This approach allows productivity to be interpreted as an individual-level characteristic rather than an artifact of publication-level aggregation. Consequently, different or multiple affiliations in a period do not change the overall productivity level in that period. The institutional research intensity variable was used only in our model approach.

We had complete bibliometric data about individual scientists: full names, information about affiliations (from publication bylines), disciplines (the modal value: to which did they refer most frequently in all their research articles and papers in conference proceedings indexed in Scopus), and their unique Scopus ID. The modal approach reduces the influence of occasional cross-disciplinary publications and reflects the dominant orientation of a scientist's research activity. We also had the year of each scientist's first publication (indicating academic age), experience at the time of publishing any article or at the end of each sex-year period. Academic age thus captures cumulative publishing exposure rather than biological age or formal career stage. To define gender, we used the gender-defining tool genderize.io, with the accuracy threshold of 0.85.

We conducted thorough data cleaning, removing from the analysis 6,885 cases for which gender could not be established, 14,431 cases for which the year of the first publication could not be established, and 49,939 cases for which the dominant discipline could not be established. Finally, at the publication level, publications with more than 100 authors were removed from the dataset. The areas of study were restricted to 15 STEMM disciplines (science, technology, engineering, mathematics, and medicine according to the All Science Journal Classification [ASJC] used in Scopus). Finally, the dataset included more than 144,000 scientists and more than 433,000 publications.

## 3.2. Sample description

The analysis is based on a full-population dataset (all internationally visible scientists) comprising $N = 144,314$ Polish researchers publishing between 1992 and 2021 whose output consisted of a total of $N_{art} = 433,546$ unique research articles indexed in Scopus. The data were aggregated into five six-year analytical periods, which allowed us to capture long-term structural changes in the science system, rather than merely short-term fluctuations. The authors came from all science sectors (higher education, research institutes, and the private sector) and could be employed full-time or part-time. For productivity calculations, we used only research articles (minimum threshold: one article in the whole period).

In all periods, an increase in the size of each academic age group is observed, with the fastest growth in the youngest cohort. The number of articles steadily increased in successive analytical periods – from 28,241 publications in 1992–1997 to 156,611 in 2016–2021. Table 1 shows the distribution of researchers in each subperiod: some scientists published in one period, others in all of them. Therefore, the total of scientists for all periods (264,283) is much higher than the number of unique scientists (144,314).

**Table 1.** Sample description – authors in the six periods

| Feature | Category | 1992–1997 | 1998–2003 | 2004–2009 | 2010–2015 | 2016–2021 | Total |
|---|---|---|---|---|---|---|---|
| Gender | female | 6,935 | 12,979 | 22,509 | 34,545 | 46,417 | 123,385 |
|  | male | 12,299 | 18,656 | 26,894 | 37,749 | 45,300 | 140,898 |
| Discipline | AGRI | 1,338 | 2,508 | 4,328 | 7,158 | 9,223 | 24,555 |
|  | BIO | 2,162 | 3,428 | 4,947 | 7,018 | 8,835 | 26,390 |
|  | CHEM | 2,767 | 4,316 | 5,714 | 7,453 | 8,157 | 28,407 |
|  | CHEMENG | 167 | 312 | 463 | 473 | 551 | 1,966 |
|  | COMP | 154 | 282 | 657 | 1,413 | 1,757 | 4,263 |
|  | EARTH | 691 | 1,292 | 1,707 | 2,510 | 3,282 | 9,482 |
|  | ENER | 34 | 80 | 215 | 537 | 1,029 | 1,895 |
|  | ENG | 811 | 1,342 | 3,051 | 5,424 | 7,788 | 18,416 |
|  | ENVI | 322 | 678 | 1,305 | 2,465 | 3,831 | 8,601 |
|  | MATER | 752 | 1,151 | 2,105 | 3,624 | 4,963 | 12,595 |
| | MATH | 856 | 1,207 | 1,633 | 2,135 | 2,208 | 8,039 |
| | MED | 5,606 | 10,381 | 17,771 | 25,374 | 32,672 | 91,804 |
| | NEURO | 202 | 335 | 419 | 619 | 871 | 2,446 |
| | PHARM | 305 | 331 | 361 | 470 | 525 | 1,992 |
| | PHYS | 3,067 | 3,992 | 4,727 | 5,621 | 6,025 | 23,432 |
| Academic age groups | 0–9 | 11,640 | 20,386 | 31,223 | 44,186 | 52,981 | 160,416 |
| | 10–19 | 4,591 | 5,836 | 10,136 | 17,183 | 22,304 | 60,050 |
| | 20–29 | 2,552 | 4,039 | 4,943 | 5,973 | 10,170 | 27,677 |
| | 30+ | 451 | 1,374 | 3,101 | 4,952 | 6,262 | 16,140 |
| Total | | 19,234 | 31,635 | 49,403 | 72,294 | 91,717 | 264,283 |

Note: The number of scientists is higher than the number of unique scientists. Each scientist can belong to 1–5 periods based on their publishing history.

**Table 2.** Sample description – articles

| Period | Number of articles |
|---|---|
| 1992–1997 | 28,241 |
| 1998–2003 | 49,869 |
| 2004–2009 | 78,547 |
| 2010–2015 | 120,278 |
| 2016–2021 | 156,611 |
| Total | 433,546 |

## 3.3. Methods

We used bibliometric publication and citation metadata from Scopus. As in recent research (Kwiek & Roszka, 2024, 2025), our general approach has four major features. It is longitudinal (we trace Polish scientists over subsequent six-year periods, spanning the thirty years from 1992 to 2021); it is relative (we allocate scientists to the two classes of top performers and the remaining scientists: the top 10% of scientists in terms of productivity within disciplines and within periods, and the remaining 90%); it is classificatory (we allocate all scientists to one of the two classes based on their productivity rank); and it is nonlinear and journal prestige-normalized (we recognize that linear and non-normalized approaches to productivity do not reflect the differences between publishing in top-tier journals and publishing in bottom-tier journals strongly enough, as global academic journals are highly stratified).

In our approach to measuring productivity, we amplified the distance between elite journals and mid-tier journals, thereby more accurately modeling the steep stratification of the publication market. We linked individual publications to journals and their global Scopus percentile ranks (0–99) using our dataset; we were not able to link our publications to their publications over time. Working with article-level citation metrics would be preferable; however, in the Polish context, working with journal percentile ranks is very close to working with publication points routinely used by Polish scientists in the past three decades. The basic assumption of Polish research evaluation systems has been that academic journals are not equal, which is expressed in the number of points attached to them (currently in the 20–200 range). The correlation of Scopus journal lists and national lists of journals (with the points assigned), for the STEM disciplines in this study, is very high.

We constructed individual publication portfolios separately for periods. The portfolios included all Scopus-derived metadata on publications: journal metadata and publication metadata. We chose all internationally visible scientists in 15 STEMM disciplines with at least one journal article indexed in the Scopus database (1992–2021). All had an unambiguously defined gender (binary; male/female, as in the Polish registry of scientists), discipline (in each period), and academic age (i.e., publishing experience). We measured productivity in six-year periods by using publications from individual publishing portfolios and a nonlinear journal prestige-normalized approach. Only the longest working scientists had their productivity calculated in all periods (30 years).

We acted as follows: for every discipline and every period, we ranked all scientists using a full-counting, journal prestige-normalized approach to productivity. Every scientist had a dominant discipline in a period. Dominant disciplines were computed using the modal value from all disciplines (ASJC) assigned to all reference lists accompanying all articles and papers in conference proceedings published by a scientist in a period. We used the ICSR Lab database (run by Elsevier for research purposes) to link all cited references from all publications to individuals. Scientists could be assigned to the same modal discipline in all whole publishing periods or to several modal disciplines in several periods. In this granular way, we avoided a more general approach in which scientists are assigned to a single discipline lifetime. We used the 2-digit (27 subject-level fields) classification instead of the 333 unique fields offered by Scopus because of the small number of observations of top performers per discipline per period. We believe our approach is granular enough to study disciplinary differences.

The gender of scientists was defined using genderize.io, a gender detection tool. We also used academic age calculated as the time passed from the first publication (any type) to any later publication. We defined dominant university affiliations in a given period (binary approach: research-intensive universities of the IDUB type, the rest) using affiliations in all bylines of all publications in a period. The IDUB is a national research excellence program for 2020–2026 for which 10 Polish universities were selected for additional funding. Finally, we studied the productivity of individuals in the context of the productivity of other individuals from the same six-year period (similar opportunities, availability of funding) and the same discipline using productivity classes (10%/90%) rather than publication numbers so that publishing increases and decreases over time did not affect the results.

# 4. Results

## 4.1. Bivariate approach

Table 3 presents the share of publications in journals from the 90–99 Scopus percentile ranks, broken down into top performers and others, separately for disciplines and for five consecutive time periods. The group of top performers was defined as the top 10% of authors in a given discipline and period in terms of publishing productivity, which is reflected in its stable share in each cell of the table. The differences between top performers and others were assessed using the Mann-Whitney test, which addresses the hypothesis of convergence of entire distributions; mean values are presented for descriptive purposes only.

In all disciplines and periods, the distribution of the share of publications in the 90–99 percentile ranks for top performers is clearly shifted upward compared to the distribution observed among the remaining authors. This advantage is systematic and (in the vast majority of cases) statistically significant. Lack of significance occurs only sporadically and primarily concerns disciplines with very small numbers in the earliest periods, where the power of the test is limited.

Although the differences between the two classes are presented in the table as mean values, their scale indicates that the top performers' advantage over the rest has clear empirical significance. Authors belonging to the top decile of productivity typically achieve publication shares in top journals that are about two to three times higher than the remaining scientists. In technical and engineering disciplines such as ENG, COMP, ENER, and ENVI, these differences are even stronger, and in the most recent period, the average share of 90–99 percentile publications in the top performer class often exceeds 0.40.

There is strong disciplinary heterogeneity. Technical and engineering disciplines are characterized by higher percentages of articles in the 90–99 percentile ranks in both classes of scientists. In areas such as MATH, MED, and CHEM, absolute values remain lower. Regardless of these structural differences, however, the relative advantage of the top performers over others in distributions remains clear in every discipline.

There was an increase in the share of publications in the most prestigious segment (and it was particularly strong in the most recent period of 2016–2021). This increase was observed for both classes of authors, but the difference in the top performers' distributions relative to rest is not weakened. This indicates the strength of mechanisms of productivity stratification over time: despite the powerful expansion of the publication system (and the growing number of articles and scientists), access to top journals remains concentrated in a small group of top performers.

**Table 3.** Percentage of publications in journals in the 90–99 Scopus journal percentile ranks, top performers vs. rest (Mann-Whitney test)

| Period | Discipline | Top performers (upper 10%) | Other scientists (90%) | %_top performers | Mean share of publications in top journals – top performers | Mean share of publications in top journals – other scientists |
|---|---|---|---|---|---|---|
| 1992–1997 | AGRI | 134 | 1,204 | 10.0 | 0.3270*** | 0.1245 |
| 1992–1997 | BIO | 217 | 1,945 | 10.0 | 0.3140*** | 0.1437 |
| 1992–1997 | CHEM | 277 | 2,490 | 10.0 | 0.1606*** | 0.0818 |
| 1992–1997 | CHEMENG | 17 | 150 | 10.2 | 0.2478** | 0.1700 |
| 1992–1997 | COMP | 16 | 138 | 10.4 | 0.5115 | 0.3780 |
| 1992–1997 | EARTH | 70 | 621 | 10.1 | 0.2232*** | 0.0884 |
| 1992–1997 | ENER | 4 | 30 | 11.8 | 0.8333 | 0.5000 |
| 1992–1997 | ENG | 82 | 729 | 10.1 | 0.6129*** | 0.3673 |
| 1992–1997 | ENVI | 33 | 289 | 10.2 | 0.3795*** | 0.2056 |
| 1992–1997 | MATER | 76 | 676 | 10.1 | 0.3619*** | 0.2834 |
| 1992–1997 | MATH | 86 | 770 | 10.0 | 0.2618*** | 0.0871 |
| 1992–1997 | MED | 560 | 5,046 | 10.0 | 0.2163*** | 0.0971 |
| 1992–1997 | NEURO | 21 | 181 | 10.4 | 0.1477*** | 0.0765 |
| 1992–1997 | PHARM | 31 | 274 | 10.2 | 0.1620*** | 0.0528 |
| 1992–1997 | PHYS | 307 | 2,760 | 10.0 | 0.2673*** | 0.1263 |
| 1998–2003 | AGRI | 251 | 2,257 | 10.0 | 0.2532*** | 0.1008 |
| 1998–2003 | BIO | 343 | 3,085 | 10.0 | 0.2659*** | 0.1554 |
| 1998–2003 | CHEM | 432 | 3,884 | 10.0 | 0.1884*** | 0.1016 |
| 1998–2003 | CHEMENG | 32 | 280 | 10.3 | 0.2413*** | 0.0926 |
| 1998–2003 | COMP | 29 | 253 | 10.3 | 0.4846*** | 0.2817 |
| 1998–2003 | EARTH | 130 | 1,162 | 10.1 | 0.2099*** | 0.0704 |
| 1998–2003 | ENER | 7 | 73 | 8.8 | 0.5675 | 0.3950 |
| 1998–2003 | ENG | 135 | 1,207 | 10.1 | 0.5041*** | 0.2790 |
| 1998–2003 | ENVI | 68 | 610 | 10.0 | 0.3674*** | 0.1465 |
| 1998–2003 | MATER | 116 | 1,035 | 10.1 | 0.3124*** | 0.2086 |
| 1998–2003 | MATH | 121 | 1,086 | 10.0 | 0.1841*** | 0.0640 |
| 1998–2003 | MED | 1,039 | 9,342 | 10.0 | 0.2186*** | 0.0972 |
| 1998–2003 | NEURO | 34 | 301 | 10.1 | 0.1148*** | 0.0875 |
| 1998–2003 | PHARM | 34 | 297 | 10.3 | 0.1692*** | 0.0748 |
| 1998–2003 | PHYS | 400 | 3,592 | 10.0 | 0.2715*** | 0.1321 |
| 2004–2009 | AGRI | 433 | 3,895 | 10.0 | 0.2745*** | 0.1020 |
| 2004–2009 | BIO | 495 | 4,452 | 10.0 | 0.2671*** | 0.1670 |
| 2004–2009 | CHEM | 572 | 5,142 | 10.0 | 0.1988*** | 0.1092 |
| 2004–2009 | CHEMENG | 47 | 416 | 10.2 | 0.2464*** | 0.0941 |
| 2004–2009 | COMP | 66 | 591 | 10.0 | 0.3292*** | 0.1571 |
| 2004–2009 | EARTH | 171 | 1,536 | 10.0 | 0.2035*** | 0.0766 |
| 2004–2009 | ENER | 22 | 193 | 10.2 | 0.5844*** | 0.1535 |
| 2004–2009 | ENG | 306 | 2,745 | 10.0 | 0.4400*** | 0.1384 |
| 2004–2009 | ENVI | 131 | 1,174 | 10.0 | 0.3810*** | 0.1153 |
| 2004–2009 | MATER | 211 | 1,894 | 10.0 | 0.3365*** | 0.1659 |
| 2004–2009 | MATH | 164 | 1,469 | 10.0 | 0.1834*** | 0.0647 |
| 2004–2009 | MED | 1,778 | 15,993 | 10.0 | 0.1949*** | 0.0864 |
| 2004–2009 | NEURO | 42 | 377 | 10.0 | 0.1301*** | 0.1014 |
| 2004–2009 | PHARM | 37 | 324 | 10.2 | 0.2359*** | 0.1047 |
| 2004–2009 | PHYS | 473 | 4,254 | 10.0 | 0.2697*** | 0.1445 |
| 2010–2015 | AGRI | 716 | 6,442 | 10.0 | 0.2448*** | 0.1006 |
| 2010–2015 | BIO | 702 | 6,316 | 10.0 | 0.2520*** | 0.1720 |
| 2010–2015 | CHEM | 746 | 6,707 | 10.0 | 0.1751*** | 0.1124 |
| 2010–2015 | CHEMENG | 48 | 425 | 10.1 | 0.3044*** | 0.1051 |
| 2010–2015 | COMP | 142 | 1,271 | 10.0 | 0.4103*** | 0.1349 |
| 2010–2015 | EARTH | 251 | 2,259 | 10.0 | 0.2510*** | 0.0977 |
| 2010–2015 | ENER | 54 | 483 | 10.1 | 0.4912*** | 0.1411 |
| 2010–2015 | ENG | 543 | 4,881 | 10.0 | 0.3403*** | 0.1107 |
| 2010–2015 | ENVI | 247 | 2,218 | 10.0 | 0.3142*** | 0.1191 |
| 2010–2015 | MATER | 363 | 3,261 | 10.0 | 0.2368*** | 0.1161 |
| 2010–2015 | MATH | 214 | 1921 | 10.0 | 0.1717*** | 0.0707 |
| 2010–2015 | MED | 2,538 | 22,836 | 10.0 | 0.1649*** | 0.0877 |
| 2010–2015 | NEURO | 62 | 557 | 10.0 | 0.1517*** | 0.0927 |
| 2010–2015 | PHARM | 47 | 423 | 10.0 | 0.2092*** | 0.1293 |
| 2010–2015 | PHYS | 563 | 5,058 | 10.0 | 0.2106*** | 0.1280 |
| 2016–2021 | AGRI | 923 | 8,300 | 10.0 | 0.3361*** | 0.1944 |
| 2016–2021 | BIO | 884 | 7,951 | 10.0 | 0.3352*** | 0.2836 |
| 2016–2021 | CHEM | 816 | 7,341 | 10.0 | 0.2511*** | 0.1648 |
| 2016–2021 | CHEMENG | 56 | 495 | 10.2 | 0.3603*** | 0.1432 |
| 2016–2021 | COMP | 176 | 1,581 | 10.0 | 0.4608*** | 0.2799 |
| 2016–2021 | EARTH | 329 | 2,953 | 10.0 | 0.3398*** | 0.1997 |
| 2016–2021 | ENER | 103 | 926 | 10.0 | 0.6035*** | 0.4307 |
| 2016–2021 | ENG | 779 | 7,009 | 10.0 | 0.4148*** | 0.2425 |
| 2016–2021 | ENVI | 384 | 3,447 | 10.0 | 0.4050*** | 0.2657 |
| 2016–2021 | MATER | 497 | 4,466 | 10.0 | 0.2528*** | 0.1534 |
| 2016–2021 | MATH | 221 | 1,987 | 10.0 | 0.2703*** | 0.1181 |
| 2016–2021 | MED | 3,268 | 29,404 | 10.0 | 0.2270*** | 0.1440 |
| 2016–2021 | NEURO | 88 | 783 | 10.1 | 0.2581*** | 0.1844 |
| 2016–2021 | PHARM | 53 | 472 | 10.1 | 0.2535*** | 0.1285 |
| 2016–2021 | PHYS | 603 | 5,422 | 10.0 | 0.2581*** | 0.1769 |

**Figure 1.** Mean percentage of publications in the 90–99 Scopus journal percentile ranks: top performers vs. other scientists (all disciplines combined), by five six-year periods (1992–2021)

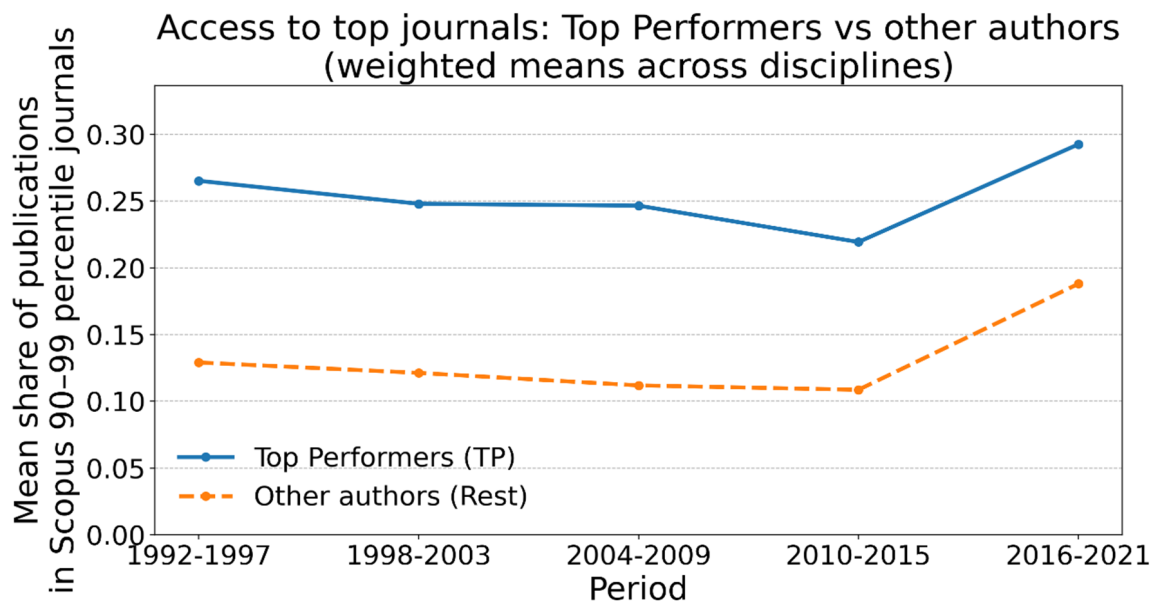

The distributions of top performers contrasted with the remaining scientists by discipline and period are shown in Figure 2. Distributions for top performers are consistently shifted toward higher values. They are characterized by higher medians and upper quantiles, while distributions for other authors are strongly concentrated close to zero.

This means that most authors outside the top performer class publish in 90–99 Scopus journals percentile ranks sporadically or not at all, and the differences between the classes become apparent primarily in the upper part of the distribution. Top performers are less diversified than scientists in in the other class, suggesting greater stability of presence in prestigious publication channels.

The results show that top performer status is a strong and persistent determinant of access to journals in the 90–99 percentile ranks. The division according to productivity is therefore qualitative and structural in nature. Its mechanisms remain stable over time despite dynamic changes in the intensity of research production.

**Figure 2.** Share of publications in journals from the 90–99 Scopus journal percentile ranks in terms of top performers (TP) vs. other scientists

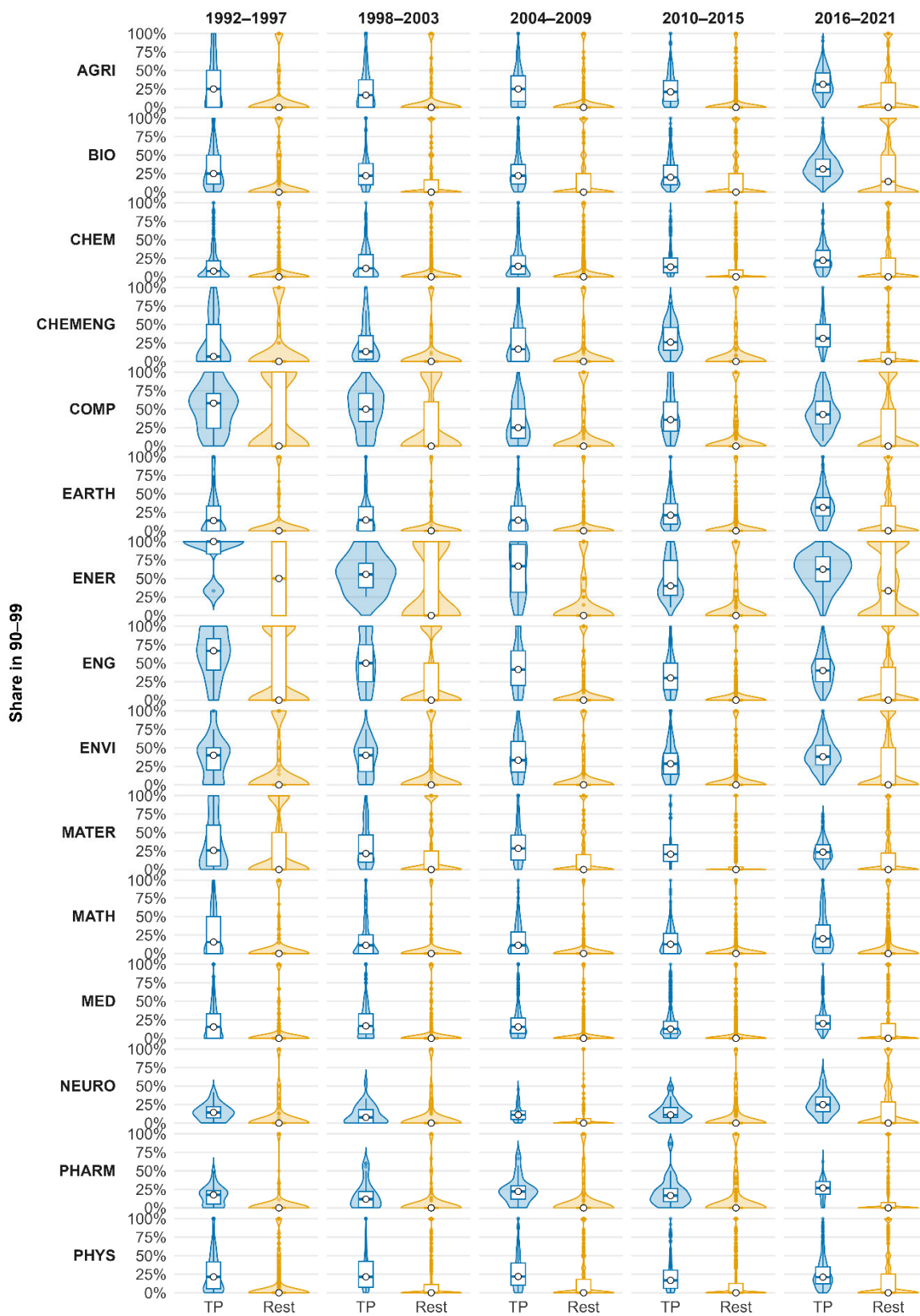

Figure 3 shows the percentages of total journal prestige-normalized production by the authors belonging to the top performer class, broken down by discipline and five consecutive time periods from 1992–1997 to 2016–2021. Each bar represents the percentage share of production generated by the top 10% of authors in a given period × discipline cell, thus illustrating the scale of production concentration measured not only by the number of publication, but also by the journal prestige. In the construction of journal prestige-normalized production, journal prestige is taken into account in a nonlinear manner: the weight of publications increases with the percentile of the journal according to the power of 2.5 (the Scopus percentile ranks ranging from zero to one), which means that works published in the highest segment of the journal ranking have a disproportionately large impact on the total value of the indicator.

**Figure 3.** Production of the 10% top-performing group in terms of nonlinear journal prestige-normalized productivity ($x^{2.5}$)

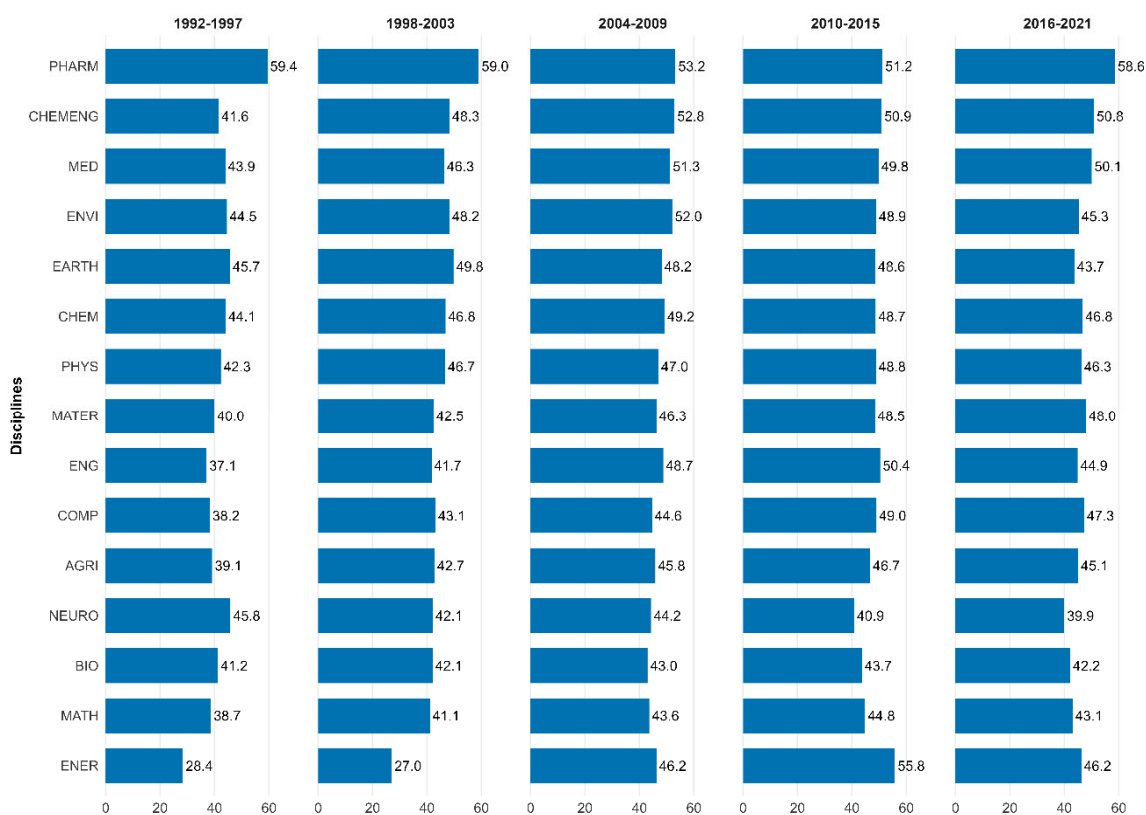

A strong concentration of research production is visible at first glance. In all disciplines and in all periods, the top decile of authors generates a significant portion of total knowledge production – usually from about one-third to more than half. This level of concentration means that production, taking into account the prestige of publication channels, is highly concentrated in a relatively narrow class of researchers (the 10/50 rule examined in Kwiek & Roszka, 2024: 10% of scientists in the Polish science system are responsible for 50% of national knowledge production). The nonlinear weighting approach to productivity means that a moderate advantage in access to more prestigious journals translates into significantly higher productivity (in a linear approach, Scopus journal percentile ranks are used according to the formula y=x, in our nonlinear approach, the rand are used according to the formula ($y = x^{2.5}$).

The scale of concentration varies between disciplines, but the pattern itself remains stable. The highest percentages of publications produced by top performers are consistently observed in areas such as PHARM, CHEMENG, MED, ENVI, and PHYS. On these disciplines, the share of production by the top decile of authors often exceeds 50%). These are also disciplines in which publishing in the highest percentile journals plays a particularly important role. In disciplines such as MATH, COMP, and AGRI, production is distributed more evenly, although even there, it remains clearly concentrated in the hands of a small proportion of authors, especially in later periods.

There has been an increase in the concentration of production from the 1990s to the mid-2000s, followed by stabilization at a high level. In some fields, including ENG, ENER, and PHYS, the share of total production by top performers increased after 2010. This suggests that the advantage of the top decile of authors not only persisted but was strengthened in some areas.

It is worth noting that even in the disciplines where the share of production by top performers is lowest, productivity is not evenly distributed. Although authors outside the top performer class generate the majority of publications in purely quantitative terms, these are largely works published in less prestigious journals. Due to the nonlinear weighting of prestige, their total contribution to productivity therefore remains clearly limited.

Overall, the concentration of research production in Polish science is structural and stable over time. Top performers both publish more and dominate the segment of highly prestigious publications, which (given the weighting method used) translates into a disproportionately large share of total production. This result complements earlier analyses and confirms that stratification by productivity is one of the key mechanisms organizing access to prestigious journals.

Figure 4 shows the distribution of articles by Scopus journal deciles (expressed as percentile ranks), broken down into the top performer class and the remaining scientists. The analysis is conducted separately for disciplines and subsequent time periods. The horizontal axis shows journal deciles (from 0–9 to 90–99), while the vertical axis shows the share of articles published in a given quality segment. The lines show the average shares for the two classes of scientists and allow for a direct comparison of their publication profiles across the entire quality scale.

In almost all disciplines and periods, a clear quality gradient is visible. The share of articles systematically increases with the transition to higher journal deciles. The highest concentration of publications appears in the three deciles of 70–79, 80–89, and 90–99. Against this background, the differences between top performers and the remaining scientists are most evident in the highest deciles. The curves for top performers are consistently shifted upward relative to other scientists in the 90–99 percentile segment. In many disciplines this difference begins to be visible in the 70–79 and 80–89 deciles. In later periods, this shift becomes more pronounced, indicating that the advantage of top performers comes from access to prestigious journals.

In the lower and middle parts of the quality distribution (up to about the 40–49 percentile), the publication profiles of both groups are very similar, and in many cases they almost overlap. This suggests that the differentiation between top performers and the rest is not about publishing "more everywhere" but only manifests itself at the top of the journal hierarchy.

Despite noticeable differences between disciplines – evident, for example, in the different publication profiles in PHYS, MED, COMP, and ENER – the relationship between top performers and the remaining scientists remains surprisingly stable. Regardless of discipline, the advantage of top performers is small in the lower deciles and clear in the highest ones, which suggests the existence of a common selection mechanism for the publishing elite.

In dynamic terms, a gradual intensification of concentration in the highest deciles is visible after 2010, especially in 2016–2021. This observation is consistent with earlier findings regarding the growing share of publications in the 90–99 percentile segment and the increasing concentration of production as measured by the journal prestige-normalized productivity index in the top performer class.

Overall, Figure 4 shows that the differences between top performers and other scientists are primarily qualitative rather than quantitative. Top performers do not publish more frequently across the entire spectrum of journals but disproportionately publish in the highest rated publication channels. At the same time, the similarity of distributions in the lower deciles indicates that the stratification of science is mainly evident at the top of the journal hierarchy, rather than in its lower segments.

Figure 5 shows the difference between article percentages between top performers and the remaining authors in successive deciles of Scopus journals. The analysis is conducted separately for disciplines and for five periods from 1992–1997 to 2016–2021. Each cell of the heat map represents the value

$$\Delta = share_{TP} - \text{share}_{Rest},$$

where darker shades indicate a top performer (*TP*) advantage, and light shades indicate advantage for the rest (crosses indicate no or too few observations).

The plot reveals a regular pattern of changes in the sign of differences along the journal quality scale. In the lower and middle deciles of journal ranks (up to about the 40–49 percentiles), differences are close to zero or negative. This means that top performers do not dominate in publishing in low-quality journals. In these segments, the publication profiles of both groups are very similar. It is different in deciles 70–79, 80–89, and 90–99: the differences become positive and often exceed 20 percentage points. This means that the top performer advantage is concentrated almost exclusively at the top of the journal hierarchy. And that the advantage manifests itself in the form of overrepresentation of publications in the most prestigious channels.

In dynamic terms, a change in the intensity of this pattern is observable. In the earliest period (1992–1997), positive values in the highest deciles are present. In subsequent periods, especially after 2010, they form distinct dark bands in the 90–99 percentile segment. This indicates a gradual strengthening of the concentration of publication quality in the top performer class.

**Figure 4.** Article average percentages in journal deciles, top performers (TP) vs. other scientists. All deciles = 100%.

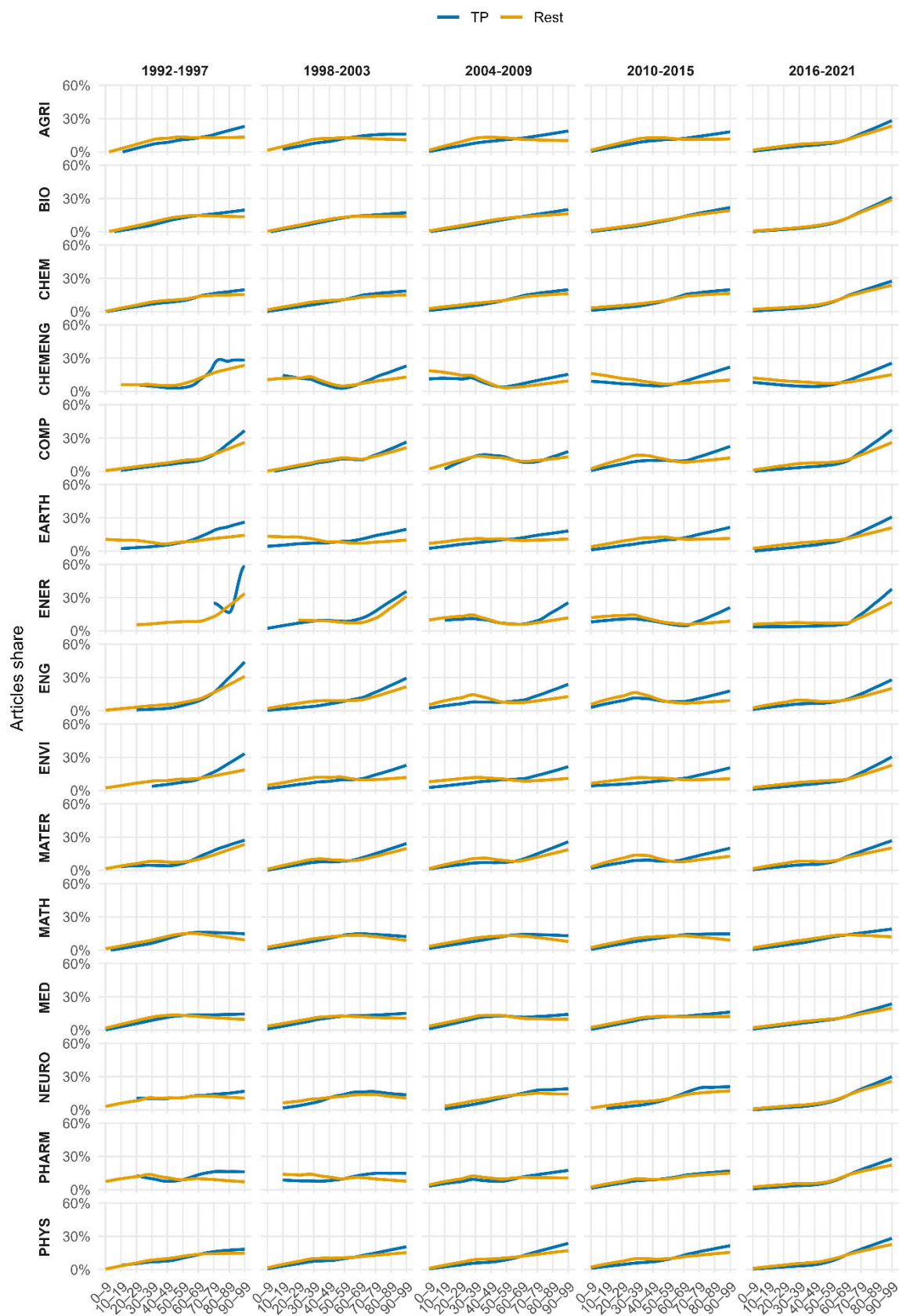

Despite clear differences between disciplines, visible, among others, in different profiles for PHYS, CHEM, MED, BIO, ENG, ENER, and COMP, the relationship between top performers and the other scientists remains surprisingly consistent. Regardless of the discipline, positive differences appear systematically only in the highest quality deciles, while in the lower segments of the distribution of journal prestige they are weak or absent.

The pattern indicates that the advantage of top performers does not lie in publishing "more everywhere" but in selective access to the most prestigious journals. The stratification of scientific production is therefore clearly qualitative in nature: the differences between top performers and other authors are revealed almost exclusively at the top of the publishing hierarchy, and their intensity has increased over time.

These findings lead to the conclusion that inequalities in science are generated primarily by two mechanisms: prestige concentration and selection into the publishing elite. They are not produced by uniform differences across the entire publication spectrum.

**Figure 5.** Difference in shares by top performers (TP) and the rest of scientists, by journal decile, period, and discipline

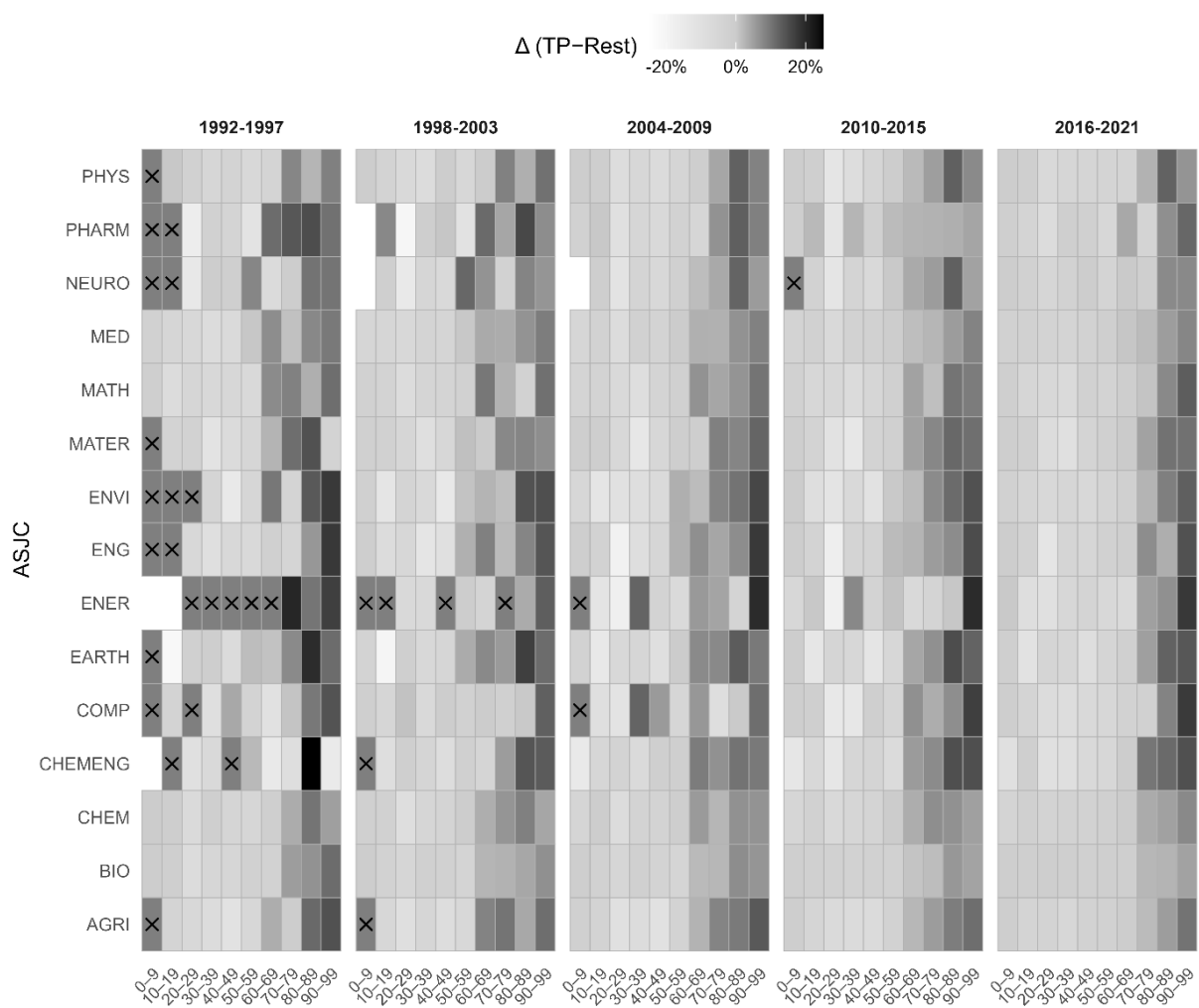

Figure 6 shows the average distribution of articles by Scopus journal decile for the top and other authors, separately for five consecutive periods, for all disciplines combined. Each curve shows the average profile of publication placement along the journal hierarchy, regardless of the specificity of individual disciplines. The shape of the distributions alone reveals a strong quality gradient.

In each period analyzed, for both top performers and other scientists, the share of articles systematically increases with the transition to higher journal deciles. This means that, on average, publications are more often placed in stronger publishing channels (which reflects the general orientation of the Polish science system toward more prestigious journals).

In the lower and middle segments of the distribution (up to about the 50–60 percentiles), the profiles of both groups are similar. The curves of the two classes run almost parallel (and in some places overlap). This suggests that publishing in low- and medium-quality journals is not the most important measure of stratification between top performers and other authors.

A clear divide only appears at the top of the journal hierarchy. In the three upper percentiles (70–79, 80–89, and especially 90–99), the top performer curve consistently separates from the curve for the rest. This difference is observable in every period and gradually increases over time. The strongest contrast is visible in the years 2016–2021, where the share of top performers' articles in the highest decile exceeds the share of the other group by several, and in some places even more than a dozen, percentage points. At the same time, it can be seen that the concentration of publications in top journals is growing for both groups, but the pace of this growth is clearly faster among top performers.

Overall, it is clear that the differences between top performers and other authors are not evenly distributed across the entire publication spectrum. They are evident almost exclusively at the top of the journal hierarchy, while in the lower and middle deciles, the distributions of both groups remain very similar. As a result, stratification in science is primarily qualitative in nature. It is driven by selective access to the most prestigious publication channels.

**Figure 6.** Article average percentages by decile (all disciplines combined)

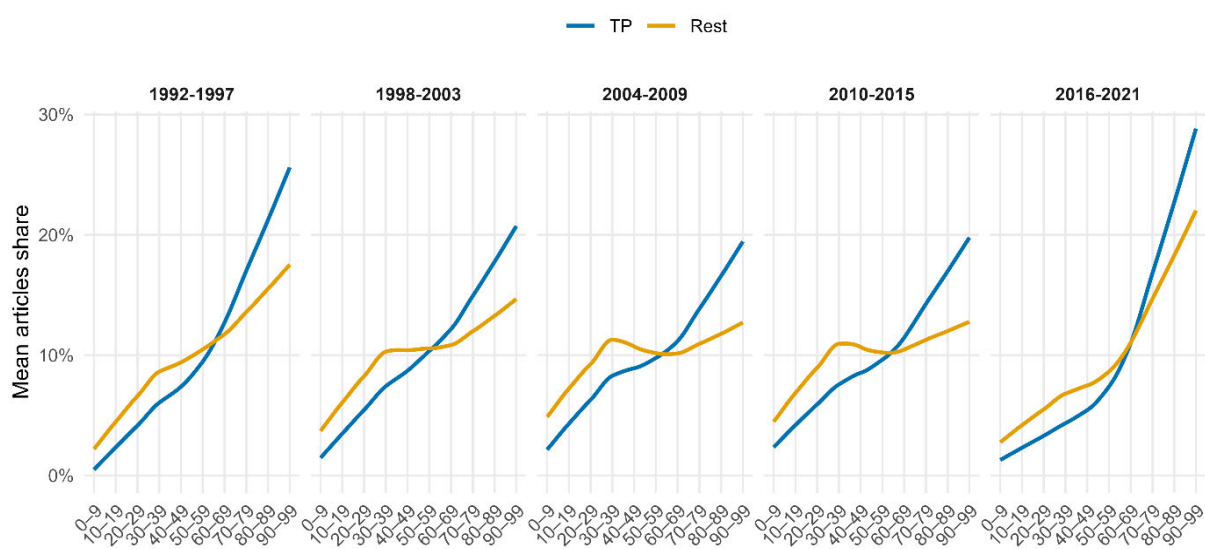

## 4.2. Model approach

The model is estimated using a micro-panel approach, where the unit of observation is a scientist active in a specific period and discipline (scientist × period × discipline). This data structure allowed us to track changes in scientists' positions over time and analyze them within comparable disciplinary contexts (see Table 6).

The dependent variable is a binary indicator of membership in the top performer class (the upper 10% in terms of journal prestige-normalized productivity). It takes the value 1 when the scientists is in the top 10% of the quality-weighted productivity distribution in a given reference cell (period × discipline), and 0 in all other cases. In other words, the model does not explain the level of productivity itself but rather the probability of entering the publishing elite under specific temporal and disciplinary conditions.

Of key importance is the construction of the journal prestige-normalized productivity measure, which forms the basis for determining top performer status by rewarding publications in the most prestigious journals and giving them disproportionately greater weight in the scientists' ranking.

The model was estimated as a generalized linear model with mixed effects (GLMM), with a function that combined a logistic function and a binomial distribution. Such a specification is natural for a binary dependent variable and allows for direct modeling of the probability of being in the top performer class, while taking into account the panel structure of the data.

The fixed part of the model was designed around two basic dimensions of scientific activity: the quantity and quality of research production. The quantitative dimension is represented by the logarithm of the number of publications by an author in a given cell (ln_pubs_c) and its quadratic term. This allowed us to capture the potential nonlinearity of the relationship. The qualitative dimension describes the mean percentile of journals in which the author published (mean_journal_pct_c). A key element of the specification is the interaction between quantity and quality (ln_pubs_c:mean_journal_pct_c), which permitted us to directly answer the question of whether high publishing intensity in average publication channels is sufficient to enter the elite of top performers, or whether locating articles in the most prestigious journals is of decisive importance.

The model also takes into account a set of demographic and institutional variables, such as gender (MaleMale), academic age (acage_c), institution type (inst_typeRest), average team size calculated on the basis of all co-authorships in a period (ATS_c), and a median prestige of publication channels, also calculated separately for each period (prestmed_c). Gender and academic age variables appear both as main effects and as moderators of the relationship between quantity and quality and the probability of entering the TOP class, which allows for testing differences in the mechanisms of selection to the publishing elite between the two classes of authors. We calculated academic age as the time passed from the first publication in the Scopus dataset (any type) to the last year of each study period. The description of independent and fixed effects variables is shown in Table 4, and the variables included in the model are presented in Table 5.

**Table 4.** Independent and fixed effects variables

| Variable (model term) | Readable name | Type/Coding |
|---|---|---|
| (Intercept) | Baseline log-odds | Constant |
| Period 1998–2003 | Period: 1998–2003 | Dummy (ref. 1992–1997) |
| Period 2004–2009 | Period: 2004–2009 | Dummy (ref. 1992–1997) |
| Period 2010–2015 | Period: 2010–2015 | Dummy (ref. 1992–1997) |
| Period 2016–2021 | Period: 2016–2021 | Dummy (ref. 1992–1997) |
| Discipline: BIO | Biology | Dummy (ref. AGRI) |

| Variable (model term) | Readable name | Type/Coding |
|---|---|---|
| Discipline: CHEM | Chemistry | Dummy (ref. AGRI) |
| Discipline: CHEMENG | Chemical Engineering | Dummy (ref. AGRI) |
| Discipline: COMP | Computer Science | Dummy (ref. AGRI) |
| Discipline: EARTH | Earth Sciences | Dummy (ref. AGRI) |
| Discipline: ENER | Energy | Dummy (ref. AGRI) |
| Discipline: ENG | Engineering | Dummy (ref. AGRI) |
| Discipline: ENVI | Environmental Sciences | Dummy (ref. AGRI) |
| Discipline: MATER | Materials Science | Dummy (ref. AGRI) |
| Discipline: MATH | Mathematics | Dummy (ref. AGRI) |
| Discipline: MED | Medicine | Dummy (ref. AGRI) |
| Discipline: NEURO | Neuroscience | Dummy (ref. AGRI) |
| Discipline: PHARM | Pharmacology | Dummy (ref. AGRI) |
| Discipline: PHYS | Physics | Dummy (ref. AGRI) |
| ln_pubs_c | Log number of publications | Continuous (centered) |
| ln_pubs2_c | Squared log publications | Continuous (centered) |
| mean_journal_pct_c | Mean journal percentile | Continuous (centered) |
| coop_c | Collaboration intensity | Continuous (centered) |
| coopint_c | International collaboration | Continuous (centered) |
| MaleMale | Male | Dummy (1 = male) |
| acage_c | Academic age | Continuous (centered) |
| inst_typeRest | Non-research-intensive institution | Dummy (ref. IDUB) |
| ATS_c | Academic time structure | Continuous (centered) |
| prestmed_c | Institutional prestige (median) | Continuous (centered) |
| ln_pubs_c:mean_journal_pct_c | Quantity × quality interaction | Interaction |

**Table 5.** Variables in the model

| Variable (model term) | Description |
|---|---|
| (Intercept) | Log-odds of being a top performer in the reference period (1992–1997), reference discipline (AGRI), and at mean values of all centered continuous variables. |
| Period 1998–2003 | Publication period indicator capturing cohort and structural time effects relative to 1992–1997. |
| Period 2004–2009 | Publication period indicator capturing cohort and structural time effects relative to 1992–1997. |
| Period 2010–2015 | Publication period indicator capturing cohort and structural time effects relative to 1992–1997. |
| Period 2016–2021 | Publication period indicator capturing cohort and structural time effects relative to 1992–1997. |
| Discipline: BIO | Discipline fixed effect capturing structural differences in publication regimes. |
| Discipline: CHEM | Discipline fixed effect capturing structural differences in publication regimes. |
| Discipline: CHEMENG | Discipline fixed effect capturing structural differences in publication regimes. |
| Discipline: COMP | Discipline fixed effect capturing structural differences in publication regimes. |
| Discipline: EARTH | Discipline fixed effect capturing structural differences in publication regimes. |
| Discipline: ENER | Discipline fixed effect capturing structural differences in publication regimes. |
| Discipline: ENG | Discipline fixed effect capturing structural differences in publication regimes. |
| Discipline: ENVI | Discipline fixed effect capturing structural differences in publication regimes. |
| Discipline: MATER | Discipline fixed effect capturing structural differences in publication regimes. |
| Discipline: MATH | Discipline fixed effect capturing structural differences in publication regimes. |
| Discipline: MED | Discipline fixed effect capturing structural differences in publication regimes. |
| Discipline: NEURO | Discipline fixed effect capturing structural differences in publication regimes. |
| Discipline: PHARM | Discipline fixed effect capturing structural differences in publication regimes. |

| Variable (model term) | Description |
|---|---|
| Discipline: PHYS | Discipline fixed effect capturing structural differences in publication regimes. |
| ln_pubs_c | Natural logarithm of the total number of publications authored in a given period, mean-centered. |
| ln_pubs2_c | Squared term of the centered log number of publications, allowing for nonlinear scale effects. |
| mean_journal_pct_c | Average percentile rank of journals in which an author publishes, capturing publication quality. |
| coop_c | Measure of co-authorship intensity, capturing the extent of collaborative publishing. |
| coopint_c | Share or intensity of international co-authorships. |
| MaleMale | Gender indicator, with female as the reference category. |
| acage_c | Years since first publication (academic career age), mean-centered. |
| inst_typeRest | Indicator for employment outside the group of research-intensive (IDUB) universities. |
| ATS_c | Measure of academic employment intensity or workload allocation. |
| prestmed_c | Median prestige of the author's institutional affiliation. |
| ln_pubs_c:mean_journal_pct_c | Interaction between publication volume and journal quality, testing whether quantity effects depend on quality. |

All quantitative variables were centered (around the mean), which facilitated the interpretation of main effects in the presence of interactions and reduced collinearity between linear and nonlinear components. In particular, it allowed main effects to refer to a scientist with an average level of other characteristics.

Structural heterogeneity is controlled by including fixed effects for period and discipline so that scientists are compared against specific historical and disciplinary conditions. At the same time, the random structure of the model includes a random author effect (1 | Auth_ID). This allows for control of unobserved, individual heterogeneity (such as permanent differences in abilities, research preferences, or network position).

**Table 6.** Statistics for the TOP 10% model

| term | Estimate | Std.Error | z.value | p.value |
|---|---|---|---|---|
| (Intercept) | -6.654 | 0.149 | -44.653 | 0.000 |
| Period 1998–2003 | -3.154 | 0.095 | -33.165 | 0.000 |
| Period 2004–2009 | -5.909 | 0.109 | -54.442 | 0.000 |
| Period 2010–2015 | -9.781 | 0.139 | -70.238 | 0.000 |
| Period 2016–2021 | -16.161 | 0.202 | -80.097 | 0.000 |
| Discipline: BIO | -2.817 | 0.101 | -27.927 | 0.000 |
| Discipline: CHEM | -7.221 | 0.129 | -55.952 | 0.000 |
| Discipline: CHEMENG | 6.654 | 0.224 | 29.640 | 0.000 |
| Discipline: COMP | 7.051 | 0.182 | 38.760 | 0.000 |
| Discipline: EARTH | 3.774 | 0.132 | 28.543 | 0.000 |
| Discipline: ENER | 8.071 | 0.256 | 31.472 | 0.000 |
| Discipline: ENG | 6.122 | 0.123 | 49.741 | 0.000 |
| Discipline: ENVI | 3.845 | 0.136 | 28.374 | 0.000 |
| Discipline: MATER | -2.607 | 0.124 | -21.027 | 0.000 |
| Discipline: MATH | 2.221 | 0.134 | 16.607 | 0.000 |
| Discipline: MED | -1.227 | 0.080 | -15.406 | 0.000 |
| Discipline: NEURO | -2.010 | 0.224 | -8.987 | 0.000 |
| Discipline: PHARM | -1.854 | 0.275 | -6.736 | 0.000 |
| Discipline: PHYS | -8.574 | 0.144 | -59.709 | 0.000 |
| ln_pubs_c | 10.843 | 0.238 | 45.601 | 0.000 |
| ln_pubs2_c | 3.762 | 0.119 | 31.585 | 0.000 |
| mean_journal_pct_c | 0.064 | 0.005 | 12.786 | 0.000 |
| coop_c | 0.517 | 0.159 | 3.254 | 0.001 |
| coopint_c | 0.087 | 0.077 | 1.123 | 0.261 |
| MaleMale | 0.012 | 0.046 | 0.266 | 0.791 |
| acage_c | 0.010 | 0.002 | 4.373 | 0.000 |
| inst_typeRest | -0.137 | 0.047 | -2.896 | 0.004 |
| ATS_c | -0.008 | 0.005 | -1.789 | 0.074 |
| prestmed_c | 0.017 | 0.003 | 5.665 | 0.000 |
| ln_pubs_c:mean_journal_pct_c | 0.185 | 0.004 | 42.086 | 0.000 |

The estimated logistic model was designed to determine what quantitative, qualitative, and structural mechanisms shape the probability of belonging to the group of top performers in science, and whether the quantity and quality of publication output act as substitute factors or instead reinforce each other.

The binary nature of the dependent variable, which identifies entry into the upper segment of the productivity distribution, justifies the use of a logit-type link function. All continuous variables were centered, and the number of publications was logarithmically transformed and expanded by a quadratic component. This facilitated the interpretation of parameters, limited the influence of extreme observations, and allowed for the capture of potential nonlinearities in the relationship.

The free term of the model takes the value $-6.654$ (SE = 0.149; z = $-44.653$; p < 0.001) and refers to a hypothetical scientist belonging to all reference categories of qualitative variables and characterized by average values of continuous variables. The high absolute negative value of this parameter indicates that the baseline probability of being in the top performer class is very low, which well reflects the publishing elite and the relative nature of the category.

The effects of time periods reveal a strong and systematic decline in the chances of achieving top performer status in subsequent time intervals relative to the reference period (1992–1997). In 1998–2003, the log-odds ratio is lower by 3.154 (SE = 0.095; z = $-33.165$; p < 0.001), in 2004–2009 by 5.909 (SE = 0.109; z = $-54.442$; p < 0.001), between 2010 and 2015 by 9.781 (SE = 0.139; z = $-70.238$; p < 0.001), and in 2016–2021 by as much as 16.161 (SE = 0.202; z = $-80.097$; p < 0.001). The monotonous increase in the absolute magnitude of these effects indicates that, with individual and structural characteristics remaining constant, entry into the publishing elite becomes increasingly difficult over time. This result suggests growing competition and relativity in selection for the publishing elite and provides a useful background for the interpretation of quantitative and qualitative effects.

The model also reveals striking cross-disciplinary differences. Compared to the reference field (AGRI), significantly lower chances of achieving top performer status are observed in, e.g., PHYS (β = $-8.574$; SE = 0.144; z = $-59.709$; p < 0.001), CHEM (β = $-7.221$; SE = 0.129; z = $-55.952$; p < 0.001), and BIO (β = $-2.817$; SE = 0.101; z = $-27.927$; p < 0.001). In turn, significantly higher odds of belonging to the publishing elite are observed in ENER, COMP, CHEMENG, and ENG, as well as in EARTH, ENVI, and MATH. The scale of these differences confirms that disciplinary context constitutes a permanent structural constraint (or support) for entry into the publishing elite, regardless of the individual scale and quality of production.

The central element of the model, directly related to our research questions, is the impact of quantitative productivity, journal quality, and their mutual interaction. The logarithm of the number of publications shows a very strong positive effect ($\beta = 10.843$; SE = 0.238; z = 45.601; p < 0.001), and a positive and significant quadratic component ($\beta = 3.762$; SE = 0.119; z = 31.585; p < 0.001) indicates this relationship is nonlinear and accelerating. This means that as quantitative productivity increases, its marginal impact on the chances of entering the publishing elite becomes stronger.

Journal quality, measured by the average Scopus percentile rank and centered, has a positive and significant effect on the log-likelihood ratio of belonging to the top performer class ($\beta = 0.064$; SE = 0.005; z = 12.786; p < 0.001), which means that publishing in more prestigious journals increases the likelihood of entering the publishing elite even with an average production scale. However, the positive and very strong interaction between the logarithm of the number of publications and the average percentile of journals ($\beta = 0.185$; SE = 0.004; z = 42.086; p < 0.001) plays a key role. Since both variables have been centered, this parameter should be interpreted as an increase in the impact of quantitative productivity on the chances of achieving top performer status as the quality of publication channels rises above the average level.

This result clearly indicates that the quantity and quality of publications operate in a complementary rather than a substitutive manner. A high number of publications significantly increases the chances of belonging to the publishing elite, especially when these publications are in high-prestige journals. On the other hand, intensifying production in low-prestige outlets brings relatively weaker selection benefits, and the high quality of individual publications alone, with a low scale of production, also does not maximize the probability of entering the class of top performers. The model thus empirically confirms that selection into the publishing elite is based on the cumulative interaction of quantity and quality.

Among the other control variables, the positive and significant effect of overall collaboration ($\beta = 0.517$; SE = 0.159; z = 3.254; p $\approx$ 0.001) indicates that scientists with a higher-than-average level of co-authorship are more likely to achieve top performer status. International collaboration does not reach statistical significance ($\beta = 0.087$; SE = 0.077; z = 1.123; p = 0.261), which suggests that its independent effect disappears when the scale and quality of production are taken into account. The lack of a significant gender effect ($\beta = 0.012$; SE = 0.046; z = 0.266; p = 0.791) indicates that, after controlling for other factors, there are no gender differences in the likelihood of belonging to the publishing elite. Academic age has a positive and significant, albeit moderate, effect ($\beta = 0.010$; SE = 0.002; z = 4.373; p < 0.001), while the negative and significant effect of institution type ($\beta = -0.137$; SE = 0.047; z = $-2.896$; p = 0.004) indicates persistent structural differences between institutional contexts. The average team size shows a weak, negative, and marginally significant effect ($\beta = -0.008$; SE = 0.005; z = $-1.789$; p = 0.074), while the median journal prestige has a positive and significant effect ($\beta = 0.017$; SE = 0.003; z = 5.665; p < 0.001), suggesting that consistently high-quality publication channels bring additional selection benefits regardless of the average quality level.

**Table 7.** Model quality statistics

| Statistic | Value |
|---|---|
| n_obs | 264283 |
| n_authors | 144314 |
| logLik | -7755.794 |
| AIC | 15573.589 |
| BIC | 15898.617 |
| is_singular | FALSE |
| RE_var_AuthID | 0.029 |
| RE_sd_AuthID | 0.171 |

| R2_marginal | 0.978 |
|---|---|
| R2_conditional | 0.978 |
| ICC_AuthID | 0.009 |
| ICC_total | 0.009 |

The model was estimated on a large sample comprising 264,283 panel observations, corresponding to 144,314 unique scientists (Table 7). This number is smaller than in the original dataset because the analysis was performed on data aggregated to the level of author × period × discipline rather than on raw publication data. In addition, only observations with complete information in all explanatory variables were used for the estimation. Cases with missing data or nonfinite values resulting from transformations (in particular logarithmic and qualitative variable constructions) were automatically excluded at the stage of constructing the database used in the estimates, which is a standard procedure in mixed model estimation. The ratio of the number of observations to the number of scientists indicates a short panel structure in which the average scientist appears in more than one but still small number of time-discipline cells. The number of scientists publishing for a long time – over all time periods – is extremely low, and our data show increases of the scientist population distributed over time periods.

The model fit was assessed based on the log-likelihood function and information criteria. The logLik value at the optimum is −7,755.79, while the AIC and BIC criteria reach 15,573.59 and 15,898.62, respectively. These measures are used for relative comparisons between alternative specifications estimated on the same sample and indicate a stable model fit given the number of parameters. It is also important that the model does not exhibit estimation singularity (is_singular = FALSE), as this means the random structure is correctly identified and the variance of the random effect has not been reduced to zero.

At the same time, the low ICC coefficient does not mean that the scientist's random effect is unnecessary. It indicates that, after taking into account the strong structural effects of period and discipline, the remaining individual heterogeneity is secondary, although still important for the correct modeling of intra-scientist relationships.

The random effect at the scientist level is characterized by a variance of 0.029, which corresponds to a standard deviation of 0.171 on the linear predictor scale. This translates into a very low intraclass correlation of approximately 0.009. In other words, after taking into account fixed effects, less than 1% of the total latent variance in the propensity to achieve top performer status can be attributed to permanent, unobservable differences between authors. At the same time, the positive value of this variance and the absence of singularities indicate that the random effect plays a corrective role, ordering intra-author relationships, but is not the main source of variation in the results.

The coefficients of determination were calculated according to the definition of Nakagawa and Schielzeth (2012), i.e., on the latent scale of the logistic model. The marginal coefficient of determination value of 0.978 indicates that almost all of the variance of the linear predictor is explained by fixed effects. A very similar value of 0.978 means that taking into account the random author effect increases the explained variance only minimally. Such a high $R^2$ value should not be interpreted in terms of the classic predictive fit known from linear regression but as a measure of the extent to which the identified structural factors determine the latent "propensity" to belong to the top performer class.

The high value is a direct consequence of the specificity of the phenomenon under study and the adopted construction of the dependent variable. Top performer status is relative and selective. Its definition is closely linked to the distributions of productivity and publication quality within periods and disciplines. At the same time, the model contains very strong structural effects of period and discipline and a clearly nonlinear component of quantitative productivity coupled with the quality of journals. As a result, a significant part of the latent predictor's variance is systematic and captured by fixed effects. Random differences between scientists play a secondary role.

Overall, the diagnostic properties of the model indicate its very good identifiability, estimation stability, and high ability to describe the mechanisms of selection to the publishing elite. The model does not exhibit a numerical problem, and the random structure is correctly estimated. The fit measures confirm that the key sources of variation in the probability of achieving top performer status are structural in nature and are adequately captured by the proposed specification. The diagnostic properties of the model confirm that the dominant source of variation in the probability of entering the publishing elite comprises structural factors – period and discipline – which provide an interpretative context for the strong, nonlinear effects of production quantity and quality.

Potential collinearity (Table 8) between explanatory variables was assessed using generalized variance inflation factors (GVIF), whereby, in accordance with the recommendations of Fox and Monette (1992), values adjusted for degrees of freedom were interpreted, that is, $\text{GVIF}^{1/(2df)}$. This approach allows for comparison of the level of collinearity between predictors with different degrees of freedom, including multicategorical variables and interactions.

**Table 8.** Collinearity statistics

| term | GVIF | Df | GVIF_adj |
|---|---|---|---|
| Period | 2.689 | 4 | 1.132 |
| ln_pubs_c | 15.139 | 1 | 3.891 |
| ln_pubs2_c | 13.375 | 1 | 3.657 |
| mean_journal_pct_c | 11.852 | 1 | 3.443 |
| coop_c | 1.233 | 1 | 1.111 |
| coopint_c | 1.266 | 1 | 1.125 |
| Male | 1.049 | 1 | 1.024 |
| acage_c | 1.066 | 1 | 1.033 |
| inst_type | 1.014 | 1 | 1.007 |
| ATS_c | 1.249 | 1 | 1.118 |
| prestmed_c | 6.206 | 1 | 2.491 |
| ln_pubs_c:mean_journal_pct_c | 8.355 | 1 | 2.891 |

The adjusted GVIF values for most variables remain very low and close to unity. This applies to period effects ($GVIF_{adj}$ = 1.132 at df = 4) as well as to control variables describing collaboration, gender, academic age, type of institution, and average team size ($GVIF_{adj}$ = in the range of 1.007–1.125). These results clearly indicate the absence of significant collinearity in this part of the specification.

Inflated but still acceptable GVIF_adj values are observed for variables describing the scale and quality of publication output and their interaction. The logarithm of the number of publications reaches $GVIF_{adj}$ = 3.891, its square component 3.657, the median percentile of journals 3.443, and the interaction of quantity and quality 2.891. A moderately increased value for median journal prestige (GVIF_adj = 2.491) reflects a partial co-occurrence of quality measures, it but remains well below conservative warning thresholds. Generally, these results indicate that the centering and construction of nonlinear components effectively reduced collinearity. The parameter estimates are stable and interpretable.

**Table 9.** Overdispersion statistics

| Statistics | Value |
|------------|----------|
| chi2 | 32124.81 |
| df_resid | 264252 |
| ratio | 0.122 |
| p_value | <0.001 |

The estimated logistic model was checked for excessive dispersion by comparing Pearson's chi-square statistic ($\chi^2$) with the number of residual degrees of freedom. In the analyzed case, the value $\chi^2 = 32,124.81$ at 264,252 degrees of freedom translates into a quotient of $\frac{\chi^2}{df} = 0.122$. Such a low value clearly indicates that the variability of the data does not exceed the level implied by the model assumptions (Table 9).

What is more, this result suggests the occurrence of so-called underdispersion, i.e., a situation in which the observed variability is lower than expected. In the context of a very large sample, a binary dependent variable, and strong structural effects included in the model (period and discipline), this result is intuitive and indicates a strong structural fit rather than a specification problem. The formal test does not provide grounds for rejecting the null hypothesis of variance consistency with the model assumptions ($p = 1$). This means that the estimated standard errors are not burdened with excessive residual volatility and statistical inference remains reliable. Consequently, there is no need for additional corrections or alternative dispersion specifications.

The distribution of Pearson residuals indicates a very good fit of the model at the observational level (Table 10). Although the extreme residual values reach $-20.919$ and $57.693$, the central part of the distribution is strongly concentrated around zero (Q1 = $-0.001$, median = 0.000, Q3 = 0.000), and the mean residual is $-0.003$ with a standard deviation of 0.349. This means that for the vast majority of observations, the differences between the observed and predicted values are minimal, and a few outliers are incidental and do not indicate a systematic misfit of the model.

**Table 10**. Pearson residuals distribution

| Statistic | Value |
|-----------------|---------|
| pearson_min | -20.919 |
| pearson_q25 | -0.001 |
| pearson_median | 0.000 |
| pearson_q75 | 0.000 |
| pearson_max | 57.693 |
| pearson_mean | -0.003 |
| pearson_sd | 0.349 |

The classification ability of the model is very high, as indicated by an AUC value of 0.9985. This reflects a near-complete separation between observations belonging to the top performer class and those outside it in the predictor space. Such a result is consistent with the presence of strong structural effects of period and discipline, as well as with a pronounced, nonlinear, and complementary interaction between the quantity and quality of publication output.

Given the relative and selective construction of the dependent variable-defined as membership in the upper segment of the productivity distribution within period-discipline cells – a very high AUC value should not be interpreted as a sign of overfitting or diagnostic deficiency. Instead, it indicates that the model accurately captures the structural mechanisms governing entry into the publishing elite under the specified institutional, temporal, and disciplinary conditions.

As robustness checks (see Electronic Supplementary Material), we estimated a set of models alternative to the main specification (a logit model with period × discipline effects and quantity–quality controls) to assess whether the conclusions are driven by assumptions about the dependence structure in the panel and the functional form of productivity. First, alongside the subject-specific approach, we estimated a population-averaged logit using a generalized estimating equation approach (GEE) with clustering at the author level, which yields inferences that are robust to the random-effects specification and relies on the classical generalized estimating equations framework for correlated data (Liang & Zeger, 1986; Zeger et al., 1988). Second, to verify that the observed nonlinearities are not artifacts of imposing a polynomial in the log number of publications, we tested a more flexible functional form (e.g., splines/smooth terms), in line with standard practice in modeling nonlinear relationships in regression settings (Harrell, 2015; Hastie et al., 2009; Wood, 2017).

In addition, in a subset of analyses, we reestimated the models using alternative productivity definitions (e.g., variants based on fractional counting and different prestige weighting) to ensure that the results are not specific to a single metric. Across all these variants, the empirical pattern remained qualitatively stable: the core quantity × quality mechanism (a positive interaction effect) retained its sign and statistical significance, and the main conclusions regarding the differences between top performers vs. others did not change such that interpretation would be affected; differences concerned primarily the level of the effects (marginal, population-averaged interpretation in GEE vs. conditional interpretation in mixed models), as is expected when comparing subject-specific and population-averaged approaches (Zeger et al., 1988).

# 5. Discussion and conclusions

The aim in our analyses was to examine how research productivity and journal prestige jointly shape access to the publication elite (understood as membership in the top performer class). The analyses were based on a large bibliometric dataset covering articles indexed in the Scopus database over 30 years (1992–2021).

The unit of analysis was the scientist publishing in a specific six-year period and a specific discipline (author × period × discipline). The dataset included 433,546 unique research articles and 144,314 unique Polish authors. The analyses distinguished between a class of top performers, defined as the top decile of authors in terms of journal prestige-normalized productivity (the upper 10%), in a given discipline and period, and the class of remaining authors (90%) in a given discipline and period. The key measure of access to the publishing elite was the author's share of publications in journals belonging to the 90-99 Scopus journal percentile ranks. This measure was interpreted as a direct indicator of presence in the most prestigious channels of scientific communication. Author productivity was measured both by the number of publications and with the nonlinear journal prestige-normalized productivity index, which gives disproportionately greater weight to publications in top journals and reduces the weight of publications in bottom-level journals.

The most important result of the model is the strong and positive effect of the number of publications, the prestige of publication channels, and their interaction. Higher publishing intensity increases the chances of belonging to the publishing elite, especially when it is coupled with publishing in top journals. Thus, productivity and channel quality do not act as substitutes but reinforce each other.

Among the control variables, the overall level of cooperation (through publications) and academic age show a positive effect, while gender does not significantly affect the probability of belonging to the publishing elite after taking into account other factors. The neutral role of gender is especially interesting in the context of numerous studies linking high research productivity with being male (Abramo et al., 2009; Kwiek, 2016) – but not surprising in Poland (Kwiek, 2018a). The diagnostic characteristics of the model confirm its stability, good identifiability, and high ability to describe the mechanisms of selection to the publishing elite.

The aim of this study was to determine whether the advantage of top performers over other scientists lies solely in their higher publication intensity, or whether it is qualitative in nature and manifests itself primarily in selective access to prestigious journals. The results clearly indicate the latter. In all the disciplines and periods examined, top performers achieve an incomparably higher share of publications in prestigious journals (90–99 Scopus journal percentile ranks), while publishing in the lower segments of the journal hierarchy does not differentiate the two classes to a similar extent.

The key finding is therefore that the stratification between top performers and other authors is not evenly distributed throughout the entire publication structure but concentrated in its narrow, highest segment. The differences between the two classes of scientists are therefore not purely quantitative (that is, top performers having more publications "everywhere") but qualitative: top performers are overrepresented precisely where rejection rates are highest and access to publication channels is most limited – in top journals.

These results allow us to reinterpret the concept of publishing elite. Traditional approaches to top performers, based on the number of publications or their total volume, suggested the existence of a continuous productivity gradient. Our analyses show, in contrast, that the most important division is at the level of the quality of the publication channel and not the intensity of knowledge production itself. Elite publishing status in science thus turns out to be a discrete and threshold phenomenon: what matters is not whether a scientist publishes a lot but whether they publish in the most prestigious journals. Top performers (the upper 10%) tend to publish in top journals (the upper 10%), which makes them the global publishing elite.

This picture is consistent with the classical theories of cumulative advantage and the Matthew effect. At the same time, it shifts the focus to institutional selection mechanisms. Prestigious journals function not only as places of knowledge transmission but also as outlets that provide international visibility, citations, and further publication opportunities. Top journals are not a neutral "mirror" of research quality but an active element of the stratified structure of the global science system.

We empirically confirm that productivity and the prestige of publication channels are complementary rather than supplementary. Our models show that publishing intensity increases the likelihood of belonging to the publishing elite primarily when it is coupled with publishing in top journals. The sheer number of publications, without the quality component, has a much weaker selection power.

This finding has an important implication suggesting selection mechanisms at work in the science system: access to prestigious publication channels powerfully differentiates positions taken by the publishing research population. The publishing elite do not seem to emerge from diligence and effort only but also from successful entries into selective global publication circuits.

Our cross-disciplinary analyses show that the stratification mechanisms described above are permanent and structural. Although the absolute level of participation in top journals varies between disciplines – reflecting different publication cultures and journal structures – the relative advantage of top performers over other scientists remains consistent across disciplines. Similarly, despite the overall increase in publications and the expansion of the Polish scientific system over time, inequalities in access to top journals are not diminishing. This means that the patterns observed are not temporary; on the contrary, with the expansion of the Polish science system, the importance of selective mechanisms may be increasing as growing competition for visibility and prestige enhances the role of publication channels as selection tools.

Our findings have direct implications for research on academic careers. If access to top journals is highly concentrated in the hands of top performers and stable over time, it should be treated as a key mechanism of career gatekeeping rather than as merely a derivative of high productivity. Publishing in prestigious journals increases visibility, strengthens reputation, and influences institutional decisions regarding tenure, promotions, and grants. In this sense, the study shifts the emphasis from the question of "who publishes more" to the question of "who gains access to prestigious channels – and with what regularity." The answer to the latter question is crucial to understanding why inequalities in science are so persistent and how publishing elites reproduce themselves.

Like any bibliometric study, this one has limitations. First, the prestige of journals was operationalized using Scopus percentile ranks, which do not exhaust all dimensions of research quality and significance. The major advantage of the Scopus percentile ranks system (CiteScore) is that it is citation-based and covers about 30,000 active journals (2026). Second, our analysis focused on article publications and did not cover other forms of knowledge production, which may have important roles in some disciplines. For this reason, we included only STEMM disciplines, as in Poland, outputs in the humanities and social sciences are still predominantly published in Polish. Third, although our results suggest strong career implications, bibliometric data alone do not allow for direct observation of institutional decisions such as promotions or grant awards. Our dataset does not cover critical variables such as departmental "climates" and "personalities" or other department-level data (research intensity, access to funding, staff structure, international collaboration, gender parity, students/staff ratio, etc.; Fox & Mohapatra, 2007; Leisyte & Dee, 2012; Tang & Horta, 2023; Turner & Mairesse, 2005). However, these limitations do not undermine the main conclusions of the study but point to the need for research combining bibliometric data with administrative (and qualitative) data, especially in the context of full academic career trajectories.

In summary, we have shown that the emergence of publishing elites in science is not a simple effect of linear accumulations of publications. The advantage of top performers over other scientists is revealed primarily in selective access to prestigious journals (rather than in a uniform increase in publishing activity in the entire hierarchy of publishing channels). Top journals emerge as playing a key stratification role. From a broader perspective, our results suggest that the contemporary science system is increasingly organized around qualitative thresholds of access rather than continuous gradients of productivity. Understanding who crosses these thresholds and why is essential for a reliable analysis of inequality in science. The micro-panel dataset used in the study (scientist × period × discipline), covering a long time horizon and containing detailed information on journal prestige and the publication intensity of individuals, enables a transition from static class comparisons to analyses of dynamic selection processes.

Finally, there are a number of directions for future research. We could model access to the segment of the most prestigious journals (90–99 Scopus percentile ranks) as events in time. It is possible to define the time from the first publication to the first publication in this segment and estimate survival models taking into account time-varying variables such as the cumulative number of publications, the average prestige of previous journals, co-authorship intensity, or the share of international collaboration. This would allow us to identify qualitative changes in publication trajectories across genders and disciplines. We could also separate *entering* from *remaining* in the publishing elite. Scientists could be assigned to states of having no publications in top journals, having an episodic presence, and maintaining a regular presence in top journals. The probabilities of transitions between these states could then be analyzed to assess the strength of elite reproduction by gender and discipline.

Another area for research could be the identification of heterogeneity within the top performer class. The data allow us to distinguish between trajectories (e.g., latent class analysis) using the number of publications, the share of publications in top journals, the variability of prestige, and the rate of change in these indicators, which make it possible to identify different strategies of entering the class of top performers. An important extension of this research would also be to take into account the nonlinear effects of academic age and cohorts entering the publication system, e.g., through interactions between journal prestige and academic age in logistic regression models. This would make it possible to determine whether the importance of top journals is greater in the early stages of an academic career and whether the window of entry to the publishing elite narrows with age.

Our study clearly shows that it is top journals (the top 10%) that make the difference for membership in the top performer class (top 10%). In the Polish context, top journals tend to be open to top performers – and they tend to be relatively, sometimes fully, closed to remaining scientists. However, on a cautionary note, the top 10% of journals will always be only 10% of them. It is impossible to expect masses of scientists to be publishing in them, as their principal feature is the scarcity of publishing space.

## Acknowledgments

We gratefully acknowledge the assistance of the International Center for the Studies of Research (ICSR) Lab, with particular gratitude to Kristy James and Alick Bird. We gratefully acknowledge the support of Dr. Łukasz Szymula with Scopus data acquisition and integration.

## Author contributions

Marek Kwiek: Conceptualization, Data curation, Formal analysis, Investigation, Methodology, Resources, Software, Validation, Writing—original draft, Writing—review & editing. Wojciech Roszka: Conceptualization, Data curation, Formal analysis, Investigation, Methodology, Software, Validation, Visualization, Writing—original draft, Writing—review & editing.

## Competing interests

The authors have no competing interests.

## Funding information

We gratefully acknowledge the support provided by the Ministry of Science (NDS grant no. NdS-II/SP/0010/2023/01).

## Data availability

We used data from Scopus, a proprietary scientometric database. For legal reasons, data from Scopus received through collaboration with the ICSR Lab (Elsevier) cannot be made openly available.

# Top performers and top journals:
# Persistent concentration in scientific publishing

**Marek Kwiek**
Center for Public Policy Studies, Adam Mickiewicz University, Poznan, Poland
kwiekm@amu.edu.pl, ORCID: orcid.org/0000-0001-7953-1063

**Wojciech Roszka**
(1) Poznan University of Economics and Business, Poznan, Poland
(2) Center for Public Policy Studies, Adam Mickiewicz University, Poznan, Poland
wojciech.roszka@ue.poznan.pl, ORCID: orcid.org/0000-0003-4383-3259

## Electronic Supplementary Materials

Table ESM 1 and the corresponding Figure ESM 1 show the share of articles published in journals in the 90-99 journal percentile rank, broken down by gender, discipline, and time period. This measure can be interpreted as the probability of publications to appear in top journals and it is a direct indicator of access to the publishing elite (a class of top performers).

A striking feature of our results is strong disciplinary heterogeneity. The differences between disciplines are much greater (descriptively) than the differences between women and men within the same discipline. Technical and engineering disciplines (e.g., ENER, ENG, COMP) are characterized by relatively high shares of 90-99 percentile rank articles, while in areas such as MATH, MED, CHEM, and PHYS, the shares of publications in top journals remain systematically lower. The patterns are stable over time (over the five periods) and indicate persistent structural differences in publication patterns between scientists from different disciplines.

Against this backdrop, gender differences in the shares are moderate. In many large disciplines, the share of 90-99 percentile articles for men is slightly shifted upwards relative to the distribution observed for women. In the table, this effect is presented using mean values, but the non-parametric Mann-Whitney test does not refer to differences in means, but to the hypothesis of convergence of entire distributions (more precisely: the absence of a systematic advantage of one distribution over another).

However, the scale of these shifts is small and it most often corresponds to differences of a few percentage points. The frequent statistical significance of the Mann-Whitney test results primarily from the large sample sizes, which increase the power of the test and allow even small differences in the position of the distributions to be detected, rather than from the existence of clear, qualitatively large gender inequalities.

An analysis of the distributions (Figure ESM 1) further indicates that intra-gender variation significantly exceeds inter-gender variation. In most disciplines, the distributions are strongly skewed and dominated by observations close to zero, which means that a significant proportion of authors, regardless of gender, publish in top journals very rarely or never. At the same time, a small group of authors achieve very high values, creating long right tails in the distributions.

Taken together, these results suggest that gender differentiates access to prestigious journals in a moderate way that is strongly dependent on the disciplinary context, while the key source of inequality remains the author's position in the productivity hierarchy and structural differences between disciplines.

**Table ESM 1.** Percentage of articles in journals from the 90-99 Scopus percentile ranks, by gender, together with the Mann-Whitney test

| Period | Discipline | Women | Men | % women | Mean share for women top performers | Mean share for men top performers |
|--------|-----------|-------|-----|---------|-------------------------------------|----------------------------------|
| 1992-1997 | AGRI | 581 | 757 | 43.4 | 0.146 | 0.144 |
| 1992-1997 | BIO | 1096 | 1066 | 50.7 | 0.156 | 0.166** |
| 1992-1997 | CHEM | 1016 | 1751 | 36.7 | 0.089 | 0.090** |
| 1992-1997 | CHEMENG | 36 | 131 | 21.6 | 0.205 | 0.170 |
| 1992-1997 | COMP | 26 | 128 | 16.9 | 0.221 | 0.427* |
| 1992-1997 | EARTH | 180 | 511 | 26.0 | 0.095 | 0.104 |
| 1992-1997 | ENER | 6 | 28 | 17.6 | 0.556 | 0.536 |
| 1992-1997 | ENG | 99 | 712 | 12.2 | 0.351 | 0.398 |
| 1992-1997 | ENVI | 126 | 196 | 39.1 | 0.268 | 0.195 |
| 1992-1997 | MATER | 217 | 535 | 28.9 | 0.246 | 0.310* |
| 1992-1997 | MATH | 159 | 697 | 18.6 | 0.088 | 0.108 |
| 1992-1997 | MED | 2525 | 3081 | 45.0 | 0.102 | 0.115*** |
| 1992-1997 | NEURO | 116 | 86 | 57.4 | 0.053 | 0.125 |
| 1992-1997 | PHARM | 155 | 150 | 50.8 | 0.086 | 0.041 |
| 1992-1997 | PHYS | 597 | 2470 | 19.5 | 0.119 | 0.146*** |
| 1998-2003 | AGRI | 1182 | 1326 | 47.1 | 0.136* | 0.099 |
| 1998-2003 | BIO | 1975 | 1453 | 57.6 | 0.165 | 0.169** |
| 1998-2003 | CHEM | 1898 | 2418 | 44.0 | 0.105 | 0.114*** |
| 1998-2003 | CHEMENG | 82 | 230 | 26.3 | 0.103 | 0.110 |
| 1998-2003 | COMP | 31 | 251 | 11.0 | 0.443 | 0.285 |
| 1998-2003 | EARTH | 400 | 892 | 31.0 | 0.076 | 0.088* |
| 1998-2003 | ENER | 16 | 64 | 20.0 | 0.328 | 0.431 |
| 1998-2003 | ENG | 186 | 1156 | 13.9 | 0.283 | 0.305 |
| 1998-2003 | ENVI | 308 | 370 | 45.4 | 0.179 | 0.160 |
| 1998-2003 | MATER | 386 | 765 | 33.5 | 0.187 | 0.235** |
| 1998-2003 | MATH | 276 | 931 | 22.9 | 0.081 | 0.075 |
| 1998-2003 | MED | 5021 | 5360 | 48.4 | 0.108 | 0.110* |
| 1998-2003 | NEURO | 200 | 135 | 59.7 | 0.078 | 0.108 |
| 1998-2003 | PHARM | 196 | 135 | 59.2 | 0.102* | 0.059 |
| 1998-2003 | PHYS | 822 | 3170 | 20.6 | 0.127 | 0.151*** |
| 2004-2009 | AGRI | 2242 | 2086 | 51.8 | 0.129 | 0.108 |
| 2004-2009 | BIO | 3060 | 1887 | 61.9 | 0.175 | 0.180*** |
| 2004-2009 | CHEM | 2820 | 2894 | 49.4 | 0.114 | 0.122*** |
| 2004-2009 | CHEMENG | 154 | 309 | 33.3 | 0.161 | 0.084 |
| 2004-2009 | COMP | 104 | 553 | 15.8 | 0.200 | 0.170 |
| 2004-2009 | EARTH | 568 | 1139 | 33.3 | 0.091 | 0.089 |
| 2004-2009 | ENER | 44 | 171 | 20.5 | 0.173 | 0.204 |
| 2004-2009 | ENG | 437 | 2614 | 14.3 | 0.154 | 0.171 |
| 2004-2009 | ENVI | 645 | 660 | 49.4 | 0.155 | 0.129 |

| 2004-2009 | MATER | 740 | 1365 | 35.2 | 0.172 | 0.189* |
|---|---|---|---|---|---|---|
| 2004-2009 | MATH | 429 | 1204 | 26.3 | 0.062 | 0.082** |
| 2004-2009 | MED | 9722 | 8049 | 54.7 | 0.093 | 0.102*** |
| 2004-2009 | NEURO | 268 | 151 | 64.0 | 0.105 | 0.102 |
| 2004-2009 | PHARM | 225 | 136 | 62.3 | 0.127 | 0.103 |
| 2004-2009 | PHYS | 1051 | 3676 | 22.2 | 0.138 | 0.162*** |
| 2010-2015 | AGRI | 3944 | 3214 | 55.1 | 0.114 | 0.116* |
| 2010-2015 | BIO | 4490 | 2528 | 64.0 | 0.175 | 0.190*** |
| 2010-2015 | CHEM | 3947 | 3506 | 53.0 | 0.115 | 0.123*** |
| 2010-2015 | CHEMENG | 166 | 307 | 35.1 | 0.112 | 0.133* |
| 2010-2015 | COMP | 210 | 1203 | 14.9 | 0.168 | 0.162 |
| 2010-2015 | EARTH | 889 | 1621 | 35.4 | 0.107 | 0.116* |
| 2010-2015 | ENER | 133 | 404 | 24.8 | 0.156 | 0.183 |
| 2010-2015 | ENG | 809 | 4615 | 14.9 | 0.159 | 0.129 |
| 2010-2015 | ENVI | 1352 | 1113 | 54.8 | 0.144 | 0.132 |
| 2010-2015 | MATER | 1387 | 2237 | 38.3 | 0.142 | 0.120 |
| 2010-2015 | MATH | 551 | 1584 | 25.8 | 0.080 | 0.081* |
| 2010-2015 | MED | 14,736 | 10,638 | 58.1 | 0.091 | 0.101*** |
| 2010-2015 | NEURO | 386 | 233 | 62.4 | 0.085 | 0.121 |
| 2010-2015 | PHARM | 311 | 159 | 66.2 | 0.124 | 0.163 |
| 2010-2015 | PHYS | 1234 | 4387 | 22.0 | 0.139 | 0.136 |
| 2016-2021 | AGRI | 5287 | 3936 | 57.3 | 0.214 | 0.201 |
| 2016-2021 | BIO | 5846 | 2989 | 66.2 | 0.289 | 0.289* |
| 2016-2021 | CHEM | 4497 | 3660 | 55.1 | 0.173*** | 0.173 |
| 2016-2021 | CHEMENG | 203 | 348 | 36.8 | 0.148 | 0.176 |
| 2016-2021 | COMP | 301 | 1456 | 17.1 | 0.348* | 0.288 |
| 2016-2021 | EARTH | 1259 | 2023 | 38.4 | 0.225 | 0.207 |
| 2016-2021 | ENER | 325 | 704 | 31.6 | 0.459 | 0.443 |
| 2016-2021 | ENG | 1516 | 6272 | 19.5 | 0.260 | 0.260 |
| 2016-2021 | ENVI | 2144 | 1687 | 56.0 | 0.279 | 0.281 |
| 2016-2021 | MATER | 2149 | 2814 | 43.3 | 0.176 | 0.154 |
| 2016-2021 | MATH | 587 | 1621 | 26.6 | 0.137 | 0.132 |
| 2016-2021 | MED | 19,972 | 12,700 | 61.1 | 0.152 | 0.153*** |
| 2016-2021 | NEURO | 532 | 339 | 61.1 | 0.189 | 0.196 |
| 2016-2021 | PHARM | 370 | 155 | 70.5 | 0.137 | 0.150 |
| 2016-2021 | PHYS | 1429 | 4596 | 23.7 | 0.175 | 0.188** |

**Figure ESM 1.** Percentage of articles in journals in the 90-99 Scopus journal percentile ranks by gender

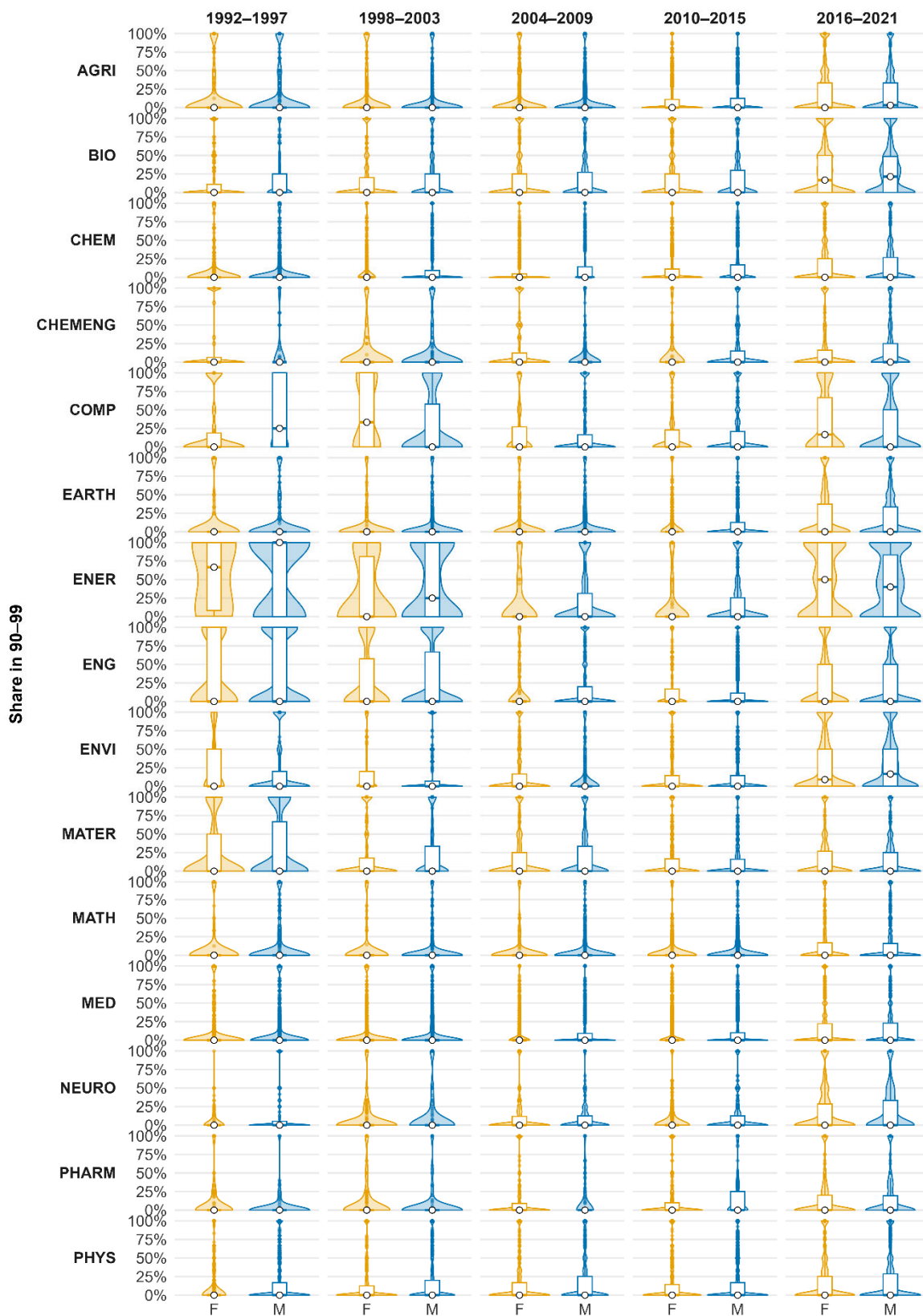

## Robustness checks

**Table ESM 2.** Results of robustness checks: Top 5% and Top 1% quality-weighted productivity models

| term | Top 5% (journal prestige-weighted productivity) | | Top 1% (Journal prestige-weighted productivity) | |
|---|---|---|---|---|
| | estimate | p.value | estimate | p.value |
| (Intercept) | -9.859 | <0.001 | -12.158 | <0.001 |
| acage_c | 0.010 | 0.001 | 0.018 | 0.001 |
| Discipline: BIO | -2.914 | <0.001 | -3.221 | <0.001 |
| Discipline: CHEM | -8.165 | <0.001 | -8.858 | <0.001 |
| Discipline: CHEMENG | 6.676 | <0.001 | 5.064 | <0.001 |
| Discipline: COMP | 7.057 | <0.001 | 4.181 | <0.001 |
| Discipline: EARTH | 3.800 | <0.001 | 3.946 | <0.001 |
| Discipline: ENER | 8.234 | <0.001 | 7.051 | <0.001 |
| Discipline: ENG | 6.207 | <0.001 | 4.835 | <0.001 |
| Discipline: ENVI | 3.813 | <0.001 | 1.981 | <0.001 |
| Discipline: MATER | -3.191 | <0.001 | -3.163 | <0.001 |
| Discipline: MATH | 2.707 | <0.001 | 3.004 | <0.001 |
| Discipline: MED | -2.082 | <0.001 | -3.577 | <0.001 |
| Discipline: NEURO | -1.653 | <0.001 | -0.839 | 0.064 |
| Discipline: PHARM | -4.739 | <0.001 | -4.992 | <0.001 |
| Discipline: PHYS | -8.927 | <0.001 | -9.182 | <0.001 |
| ATS_c | -0.004 | 0.509 | -0.002 | 0.863 |
| coop_c | 0.793 | 0.001 | 0.802 | 0.121 |
| coopint_c | 0.163 | 0.125 | -0.062 | 0.784 |
| inst_typeRest | -0.045 | 0.465 | -0.042 | 0.715 |
| ln_pubs_c | 8.732 | <0.001 | 4.653 | <0.001 |
| ln_pubs_c:mean_journal_pct_c | 0.166 | <0.001 | 0.131 | <0.001 |
| ln_pubs2_c | 3.676 | <0.001 | 3.040 | <0.001 |
| MaleMale | 0.076 | 0.195 | -0.048 | 0.682 |
| mean_journal_pct_c | 0.020 | 0.013 | -0.041 | 0.009 |
| Period1998-2003 | -3.604 | <0.001 | -3.679 | <0.001 |
| Period2004-2009 | -6.664 | <0.001 | -7.433 | <0.001 |
| Period2010-2015 | -10.590 | <0.001 | -10.483 | <0.001 |
| Period2016-2021 | -17.091 | <0.001 | -16.537 | <0.001 |
| prestmed_c | 0.026 | <0.001 | 0.032 | <0.001 |

**Table ESM 3.** Robustness checks using population-averaged logit models (GEE)

| term | Top 20% FC_PW (population-averaged logit, GEE) | | Top 10% FC_PW (population-averaged logit, GEE) | |
|---|---|---|---|---|
| | estimate | p.value | estimate | p.value |
| (Intercept) | -2.580 | <0.001 | -6.657 | <0.001 |
| acage_c | 0.007 | <0.001 | 0.010 | <0.001 |
| Discipline: BIO | -3.125 | <0.001 | -2.809 | <0.001 |
| Discipline: CHEM | -6.235 | <0.001 | -7.212 | <0.001 |
| Discipline: CHEMENG | 6.783 | <0.001 | 6.648 | <0.001 |
| Discipline: COMP | 6.939 | <0.001 | 7.074 | <0.001 |
| Discipline: EARTH | 3.622 | <0.001 | 3.780 | <0.001 |
| Discipline: ENER | 8.359 | <0.001 | 8.075 | <0.001 |
| Discipline: ENG | 5.613 | <0.001 | 6.130 | <0.001 |
| Discipline: ENVI | 3.591 | <0.001 | 3.849 | <0.001 |
| Discipline: MATER | -2.491 | <0.001 | -2.600 | <0.001 |
| Discipline: MATH | 1.806 | <0.001 | 2.219 | <0.001 |
| Discipline: MED | -0.068 | 0.432 | -1.219 | <0.001 |
| Discipline: NEURO | -2.272 | <0.001 | -1.989 | <0.001 |
| Discipline: PHARM | 1.377 | <0.001 | -1.842 | <0.001 |
| Discipline: PHYS | -8.021 | <0.001 | -8.563 | <0.001 |
| ATS_c | 0.003 | 0.390 | -0.008 | 0.053 |
| coop_c | 0.112 | 0.336 | 0.511 | 0.004 |
| coopint_c | -0.105 | 0.065 | 0.086 | 0.267 |
| inst_typeRest | 0.028 | 0.465 | -0.137 | 0.003 |
| ln_pubs_c | 14.013 | <0.001 | 10.834 | <0.001 |
| ln_pubs_c:mean_journal_pct_c | 0.205 | <0.001 | 0.185 | <0.001 |
| ln_pubs2_c | 4.238 | <0.001 | 3.765 | <0.001 |
| MaleMale | 0.041 | 0.256 | 0.012 | 0.787 |
| mean_journal_pct_c | 0.144 | <0.001 | 0.064 | <0.001 |
| Period1998-2003 | -2.535 | <0.001 | -3.151 | <0.001 |
| Period2004-2009 | -4.851 | <0.001 | -5.907 | <0.001 |
| Period2010-2015 | -8.344 | <0.001 | -9.777 | <0.001 |
| Period2016-2021 | -14.709 | <0.001 | -16.156 | <0.001 |
| prestmed_c | 0.006 | 0.030 | 0.017 | <0.001 |

**Table ESM 4.** Robustness checks using spline specifications for productivity

| term | Top 20% (quality-weighted productivity, spline) | | Top 10% (quality-weighted productivity, spline) | |
|---|---|---|---|---|
| | Estimate | p.value | Estimate | p.value |
| (Intercept) | -26.111 | 0.000 | -203.165 | 0.000 |
| acage_c | 0.007 | 0.000 | 0.010 | 0.000 |
| Discipline: BIO | -3.070 | 0.000 | -2.699 | 0.000 |
| Discipline: CHEM | -6.114 | 0.000 | -6.970 | 0.000 |
| Discipline: CHEMENG | 7.090 | 0.000 | 6.863 | 0.000 |
| Discipline: COMP | 6.964 | 0.000 | 7.005 | 0.000 |
| Discipline: EARTH | 3.617 | 0.000 | 3.732 | 0.000 |
| Discipline: ENER | 8.512 | 0.000 | 8.000 | 0.000 |
| Discipline: ENG | 5.707 | 0.000 | 6.189 | 0.000 |
| Discipline: ENVI | 3.555 | 0.000 | 3.739 | 0.000 |
| Discipline: MATER | -2.526 | 0.000 | -2.605 | 0.000 |
| Discipline: MATH | 1.767 | 0.000 | 2.189 | 0.000 |
| Discipline: MED | -0.070 | 0.264 | -1.150 | 0.000 |
| Discipline: NEURO | -2.210 | 0.000 | -1.885 | 0.000 |
| Discipline: PHARM | 1.239 | 0.000 | -1.731 | 0.000 |
| Discipline: PHYS | -7.887 | 0.000 | -8.324 | 0.000 |
| ATS_c | 0.005 | 0.208 | -0.006 | 0.179 |
| coop_c | 0.093 | 0.385 | 0.439 | 0.006 |
| coopint_c | -0.110 | 0.053 | 0.078 | 0.312 |
| inst_typeRest | 0.038 | 0.319 | -0.153 | 0.001 |
| MaleMale | 0.041 | 0.253 | 0.001 | 0.979 |
| mean_journal_pct_c | 0.187 | 0.000 | 0.087 | 0.000 |
| ns(ln_pubs_c, df = 3)1 | 26.715 | 0.000 | 130.476 | 0.000 |
| ns(ln_pubs_c, df = 3)2 | 116.036 | 0.000 | 418.678 | 0.000 |
| ns(ln_pubs_c, df = 3)3 | 143.167 | 0.000 | 215.395 | 0.000 |
| Period1998-2003 | -2.574 | 0.000 | -3.258 | 0.000 |
| Period2004-2009 | -4.874 | 0.000 | -5.969 | 0.000 |
| Period2010-2015 | -8.325 | 0.000 | -9.749 | 0.000 |
| Period2016-2021 | -14.573 | 0.000 | -15.942 | 0.000 |
| prestmed_c | 0.006 | 0.014 | 0.017 | 0.000 |